\documentclass[useAMS,usenatbib]{mn2e_astroph}

\usepackage{graphicx}
\usepackage{fancyvrb}

\def\nct#1{\nocite{#1}}

\def\dox{\delta_{\rm ox}}
\def\pmx{\psi_{\rm mx}}

\def\lag{\Delta\phi}

\def\ejx{{\cal E}_1^x}
\def\ejy{{\cal E}_1^y}
\def\edx{{\cal E}_2^x}
\def\edy{{\cal E}_2^y}

\title[Polarization distortions in PSR J1456$-$6843]
{Circular polarization in radio pulsar PSR
B1451$-$68: coherent mode transitions and intrabeam interference
}

\author[J.~Dyks, P.~Weltevrede, et al.]
{J.~Dyks$^1$, P. Weltevrede$^2$ and C.~Ilie$^2$
\\
$^1$Nicolaus Copernicus Astronomical Center, Polish Academy of Sciences, Rabia\'nska 8, 87-100, Toru\'n,
Poland\\
$^2$Jodrell Bank Centre for Astrophysics, The University of Manchester, Alan Turing Building, Manchester, M13 9PL, UK
}
\begin{document}

\date{Accepted 2020 November 30. Received 2020 October 28; in original form 2020 July 27}


\maketitle

\label{firstpage}

\begin{abstract} 
The radio emission of pulsar B1451$-$68 contains two polarization modes of
similar strength, which produce two clear orthogonal polarization angle
tracks. When viewed on a Poincar\'e sphere, the emission is composed of  
two flux patches that rotate meridionally as function of pulse longitude and pass through the Stokes V
poles, which results in transitions between orthogonal polarization modes (OPMs).
Moreover, the ratio of power in the patches is inversed once within the profile window.
It is shown that the meridional circularization is caused by a coherent OPM 
transition (COMT) produced by a varying mode ratio at a fixed quarter-wave
phase lag. 
The COMTs may be ubiquitous and difficult to detect in radio pulsar data,
because they can leave no trace in polarized fractions and they are described by equation similar
to the rotating vector model. 
The circularization, which coincides with flux minima at lower frequency, 
requires that profile components are formed by radiation 
with an oscillation phase which increases with longitude in steps of $90^\circ$
per component.
The properties can be understood as an interference pattern involving two pairs of linear orthogonal modes (or
two nonorthogonal elliptic waves). The frequency-dependent coherent superposition
of coplanar oscillations can produce the minima in the pulse profile, and thereby the
illusion of components as separate entities. The orthogonally polarized signal which is
left after such negative interference explains the enhancement of polarization degree
which is commonly observed in the minima between profile components.
\end{abstract}


\def\lap{\hbox{\hspace{4.3mm}}
         \raise1.5pt \vbox{\moveleft9pt\hbox{$<$}}
         \lower1.5pt \vbox{\moveleft9pt\hbox{$\sim$ }}
         \hbox{\hskip 0.02mm}}

\def\rwobs{R_W}
\def\rwcon{R_W}
\def\rwstr{R_W}
\def\winobs{W_{\rm in}}
\def\woutobs{W_{\rm out}}
\def\phm{\phi_m}
\def\phmi{\phi_{m, i}}
\def\thm{\theta_m}
\def\dres{\Delta\phi_{\rm res}}
\def\win{W_{\rm in}}
\def\wout{W_{\rm out}}
\def\rin{\rho_{\rm in}}
\def\rout{\rho_{\rm out}}
\def\phin{\phi_{\rm in}}
\def\phout{\phi_{\rm out}}
\def\xin{x_{\rm in}}
\def\xout{x_{\rm out}}

\def\thmin{\theta_{\rm min}^{\thinspace m}}
\def\thmax{\theta_{\rm max}^{\thinspace m}}

\section{Introduction}
\label{intro}

Studies of pulsar polarization have a half century long history, with
publications addressing many different aspects: the appearance of observed
average polarization at different frequencies (Hankins \& Rankin 2010,
Karastergiou \& Johnston 2006,
Dai et al.~2015, Noutsos et al.~2015), 
\nct{hr10, kj06, dhm15, nsk15}
various statistical and instrumental effects
(McKinnon \& Stinebring 1998), \nct{ms98} propagation effects in the interstellar medium (Karastergiou
2009), \nct{k2009} single pulse polarization (Smith et al.~2013; Mitra et al.~2015),
\nct{srm13, mar2015} and
others. Theoretical works involved recognition of basic magnetospheric geometry (RVM
ie.~the rotating vector model, Radhakrishnan \& Cooke
1969; Komesaroff 1970), \nct{rc69, k70} polarization of curvature radiation (Michel
1991, Gangadhara 2010), \nct{m91, g10} calculations of allowed magnetospheric propagation modes
(Melrose \& Stoneham 1977; Arons \& Barnard 1986), \nct{ab86, ms77} 
analytical modelling of propagation effects (Petrova \& Lyubarskii 2000),
\nct{pl00} numerical ray tracing with propagation effects (Wang et
al.~2010, Hakobyan et al.~2017), \nct{wlh10, hbp17} and empirical modelling of circular
polarization. The latter was either based on coherent mode superposition  
(Kennett \& Melrose 1998; Edwards \& Stappers 2004) \nct{km98, es04} or on the noncoherent superposition of
modes (Melrose et al.~2006). \nct{mmk2006} 

In this paper we study the variations of circular polarization in PSR B1451$-$68 (PSR J1456$-$6843)
and the associated flow of radiative power between orthogonal tracks of polarization angle (PA). 
The transition of radiative power between the orthogonal PA tracks
naturally occurs in a model which assumes that the
observed polarization state (ie.~a point or patch on the Poincar\'e sphere) 
results from a coherent superposition of orthogonally and linearly
polarized\footnote{Whenever convenient, the terms `linearly polarized',
`circularly polarized' or `elliptically polarized' will be shortened below to
`linear', `circular' and `elliptical'.} 
proper mode waves. As described in Dyks (2017, hereafter D17), \nct{d2017} 
whenever the waves have similar amplitudes and combine at
phase lag of $\sim\negthinspace90^\circ$, the orthogonal transitions occur in coincidence with
maxima of circular polarization (see figs.~11, 12, and 13
therein). This can happen in various physical situations: (1) when the
phase lag distribution keeps extending up to $90^\circ$ and the amplitude ratio is
changing with pulse longitude $\Phi$ (fig.~11 in D17); (2) when similar-amplitude waves 
are coherently summed at a longitude-dependent phase lag
(fig.~12 and 13 in D17); (3) when both the amplitude and lag are
longitude-dependent (fig.~23 in Dyks 2019, hereafter D19). 
It has been found that identification of a 
 single strong PA track with a single RVM curve 
is in general meaningless (sect.~4.3.1 in D17). 
Indeed, some observations prove 
that the RVM tracks are hard to identify without earlier recognition
of polarization modes (Mitra \& Rankin 2008).

Transition through the state of quarter-cycle lag at similar wave
amplitudes  has a simple representation on the Poincar\'e sphere, where it corresponds to
the passage near or through the V pole. 
The passage occurs during a near-meridional rotation of polarization state
on the Poincar\'e sphere. The earliest illustration of this
phenomenon can be found in the PhD thesis of Ilie (2019) \nct{i19} where the V pole
passage (VPP) can be seen directly on Poincar\'e sphere maps (B1451$-$68,
Fig.~3.45 therein). Ilie's thesis also presents distributions of ellipticity
angle $\kappa = \frac{1}{2}\arctan(V/L)$, where $V$ and $L$ are the observed
circularly and linearly polarized flux, respectively. 
This makes it possible  to recognize the VPP effect
even without the help of Poincar\'e sphere projections (eg.~PSR J1900$-$2600,
Fig.~3.64 in Ilie 2019). \nct{i19} 

Whatever is the physical reason, the passage of polarization state through the V pole 
is a transition to an opposite azimuth on the Poincar\'e sphere, and
the opposite azimuth means orthogonal PA (see figs.~2 and 3 in Dyks 2020, hereafter D20).
\nct{d2020} 
Moreover, because of its
nearpolar nature, when mapped on the traditional longitude-PA diagram, the VPP phenomenon 
produces several polarization artifacts that result from the usual
cartographic problems of circumpolar mapping (D20). For example,
PA track bifurcations and vertical spread of PA result from immersion of
the V pole within the modal patch. Excessive span of PA observed on the 
equatorward side of star's magnetic pole can also result from the traverse of flux patch
near the V pole of Poincar\'e sphere.
 
In this paper we analyse Ilie's data on PSR B1451$-$68 in more detail
(Section \ref{poin}). Then we derive analytical formulae that allow us to
describe arbitrary rotations of the polarization state on the Poincar\'e sphere. 
 In Sect.~\ref{scomt} the meridional circularization in B1451$-$68 is interpreted
as a coherent orthogonal polarization mode transition. 
In Sect.~\ref{numerical} the VPP OPM transitions are calculated numerically to present the
mapping from the Poincar\'e sphere in the case of similar strength of both modes. Some   
of the model results are identified in the polarization of B1451$-$68.
A possible geometric origin for the mode strength imbalance between the
leading and trailing side of the profile is discussed
in Sect.~\ref{leadtrail}. Then we try to use the symmetry properties 
to constrain possible physical interpretations 
of polarization observed in B1451$-$68 (Sect.~\ref{physical}). This leads to a model of pulsar emission
which is based on four-mode interference (Sect.~\ref{sinterf}).

\section{Observations and polarization characteristics}
\label{poin}

\subsection{Observations}
\label{obs}

\begin{figure}
\includegraphics[width=0.75\textwidth, angle=-90]{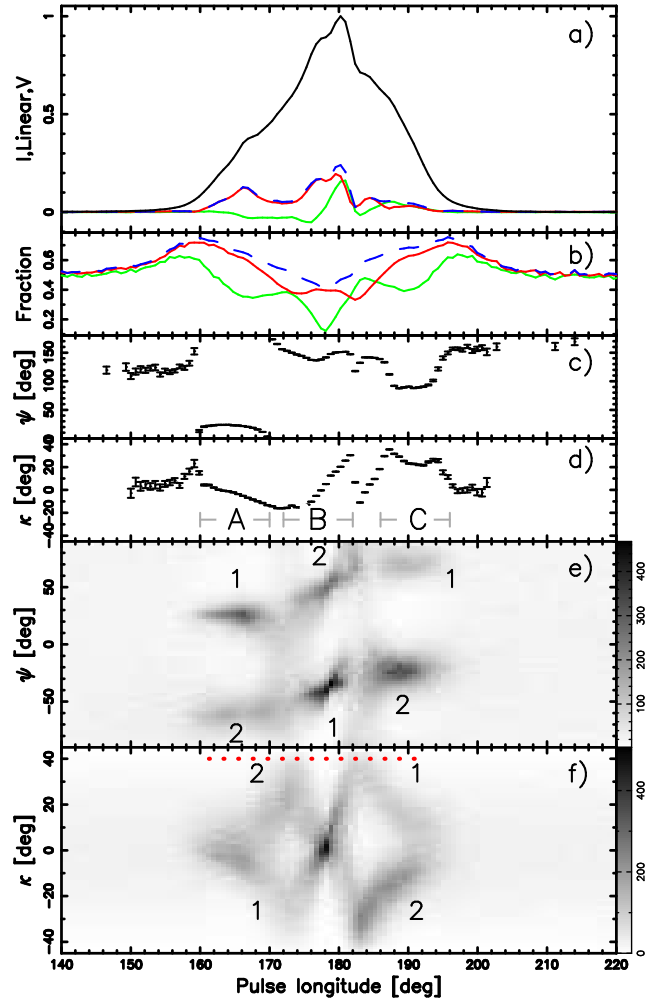}
\caption{Polarization of PSR B1451$-$68 at 1369 MHz (after Ilie 2019). 
{\bf a)} Average profile with total (black), linear (red), circular (green) and total (blue dashed) polarization.
{\bf b)} Arithmetic average of instantaneous polarized fractions: total (dashed), 
linear $L/I$ (red) and circular $|V|/I$ (green). 
Only samples for which the total polarization, $L$ and $V$ exceeds the offpulse r.m.s. are included in the average. 
{\bf c)} 
Average PA. {\bf d)} Average ellipticity angle. {\bf e)} Histogram of PA
measured in single samples for which $L$ exceeds the offpulse r.m.s. {\bf f)} Histogram of ellipticity angle measured
in single samples for which the total polarization exceeds the offpulse r.m.s. The numbers in {\bf e)} and {\bf f)} refer to different OPMs (different patches 
on the Poincar\'e sphere shown in figure \ref{data2}). Three regions of interest are labeled in {\bf d)}. The dots in panel {\bf f)} refer to the pulse longitudes explored in more detail in figure \ref{data2}.
}
\label{data1}
\end{figure}

PSR B1451$-$68 (J1456$-$6843) 
 was observed in 2016 with Parkes telescope within
a survey of modulated
radio pulsar polarization (Chapter 3 of Ilie 2019). \nct{i19} 
In total 8755 consecutive single pulses were recorded at a central frequency of 1369 MHz
with a bandwidth of 256 MHz.  
The acquired polarization data on PSR B1451$-$68 are analysed using {\sc PSRSALSA}\footnote{https://github.com/weltevrede/psrsalsa} 
(Weltevrede 2016) \nct{w2016} 
and are shown in Fig.~\ref{data1}
(reproduced Fig.~3.44 of Ilie 2019). 
Panel {\bf e)} presents the greyscale distribution of PA as measured in
single samples as a function of pulse longitude $\Phi$. Two similarly strong PA tracks 
that follow the rotating vector model (RVM) are visible. 
Both these near orthogonal PA tracks (upper and lower) consist of three sections marked A, B, and C. The
sections are separated by
breaks, ie.~longitude intervals where the PA is spread vertically over a large range of
values, hence the greyscale histogram becomes pale there.

The ellipticity angle $\kappa$, shown in the bottom panel,
forms a bow tie shape. The radiation of both modes is linearly polarized on
the leading side of the profile ($\kappa \sim 0$, $\Phi=160^\circ$), then $\kappa$ approaches
$\pm 45^\circ$, crosses zero in the middle of the profile (longitude
$178^\circ$) and again reaches
$\pm45^\circ$ before becoming nearly linear on the trailing side.

\subsection{Poincar\'e sphere view}
\label{patches}

\begin{figure*}
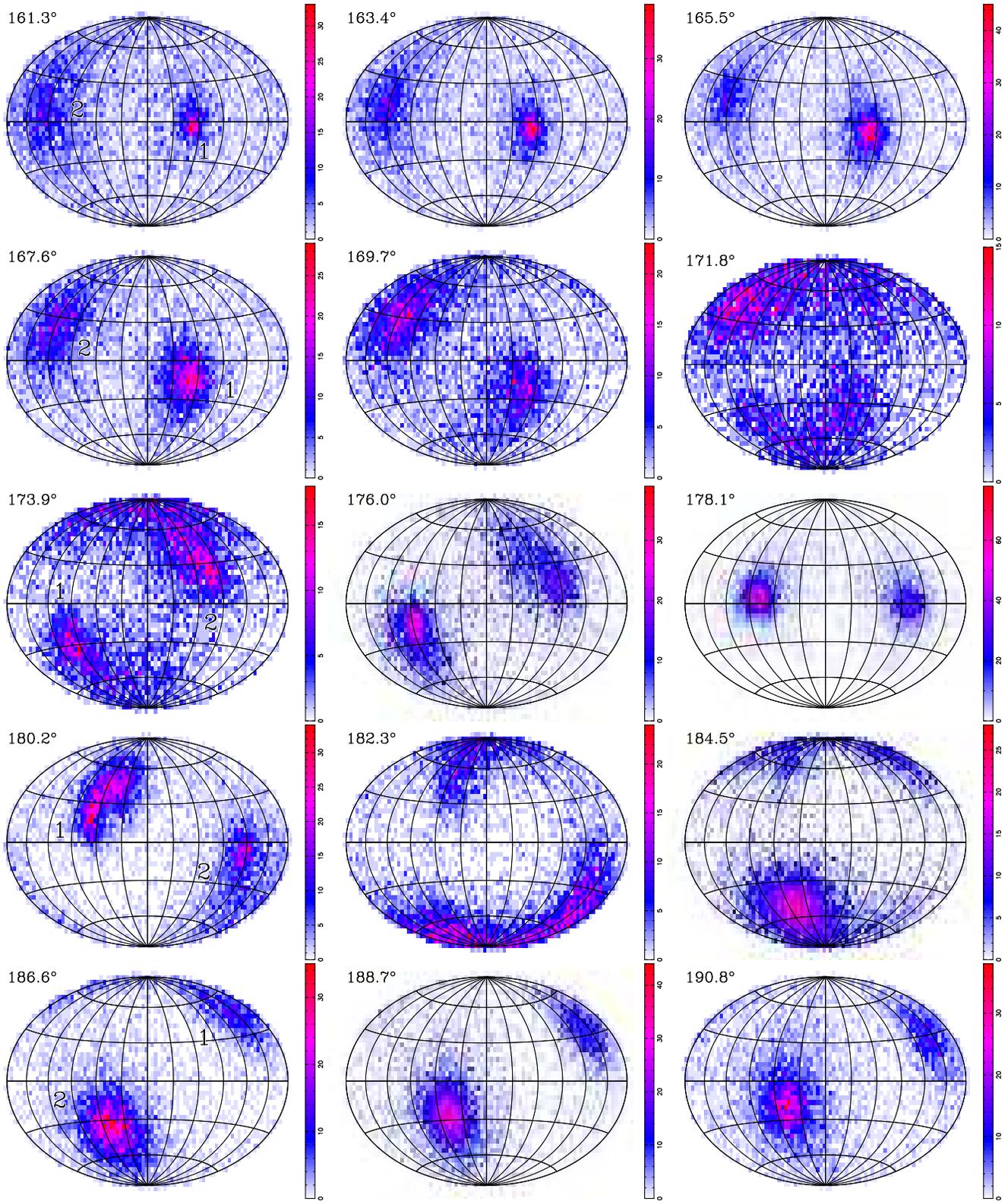

\begin{minipage}[t]{\textwidth}
\includegraphics[height=0.33\textwidth, angle=-90]{poicare_0sigma144.ps}
\includegraphics[height=0.33\textwidth, angle=-90]{poicare_0sigma147.ps}
\includegraphics[height=0.33\textwidth, angle=-90]{poicare_0sigma150.ps}

\includegraphics[height=0.33\textwidth, angle=-90]{poicare_0sigma153.ps}
\includegraphics[height=0.33\textwidth, angle=-90]{poicare_0sigma156.ps}
\includegraphics[height=0.33\textwidth, angle=-90]{poicare_0sigma159.ps}

\includegraphics[height=0.33\textwidth, angle=-90]{poicare_0sigma162.ps}
\includegraphics[height=0.33\textwidth, angle=-90]{poicare_0sigma165.ps}
\includegraphics[height=0.33\textwidth, angle=-90]{poicare_0sigma168.ps}

\includegraphics[height=0.33\textwidth, angle=-90]{poicare_0sigma171.ps}
\includegraphics[height=0.33\textwidth, angle=-90]{poicare_0sigma174.ps}
\includegraphics[height=0.33\textwidth, angle=-90]{poicare_0sigma177.ps}

\includegraphics[height=0.33\textwidth, angle=-90]{poicare_0sigma180.ps}
\includegraphics[height=0.33\textwidth, angle=-90]{poicare_0sigma183.ps}
\includegraphics[height=0.33\textwidth, angle=-90]{poicare_0sigma186.ps}
\end{minipage}
\caption{Histograms of polarization states observed at the longitudes shown in the top-left corner 
of each panel and marked with red dots in Fig.~\ref{data1}f.  
Each map presents Hammer equal area projection of Poincar\'e
sphere. In the left-hand panels the model patches are labeled consistently with the bottom panels of Fig.~\ref{data1}. 
}
\label{data2}
\end{figure*}

\begin{figure}
\includegraphics[width=0.4\textwidth, angle=-90]{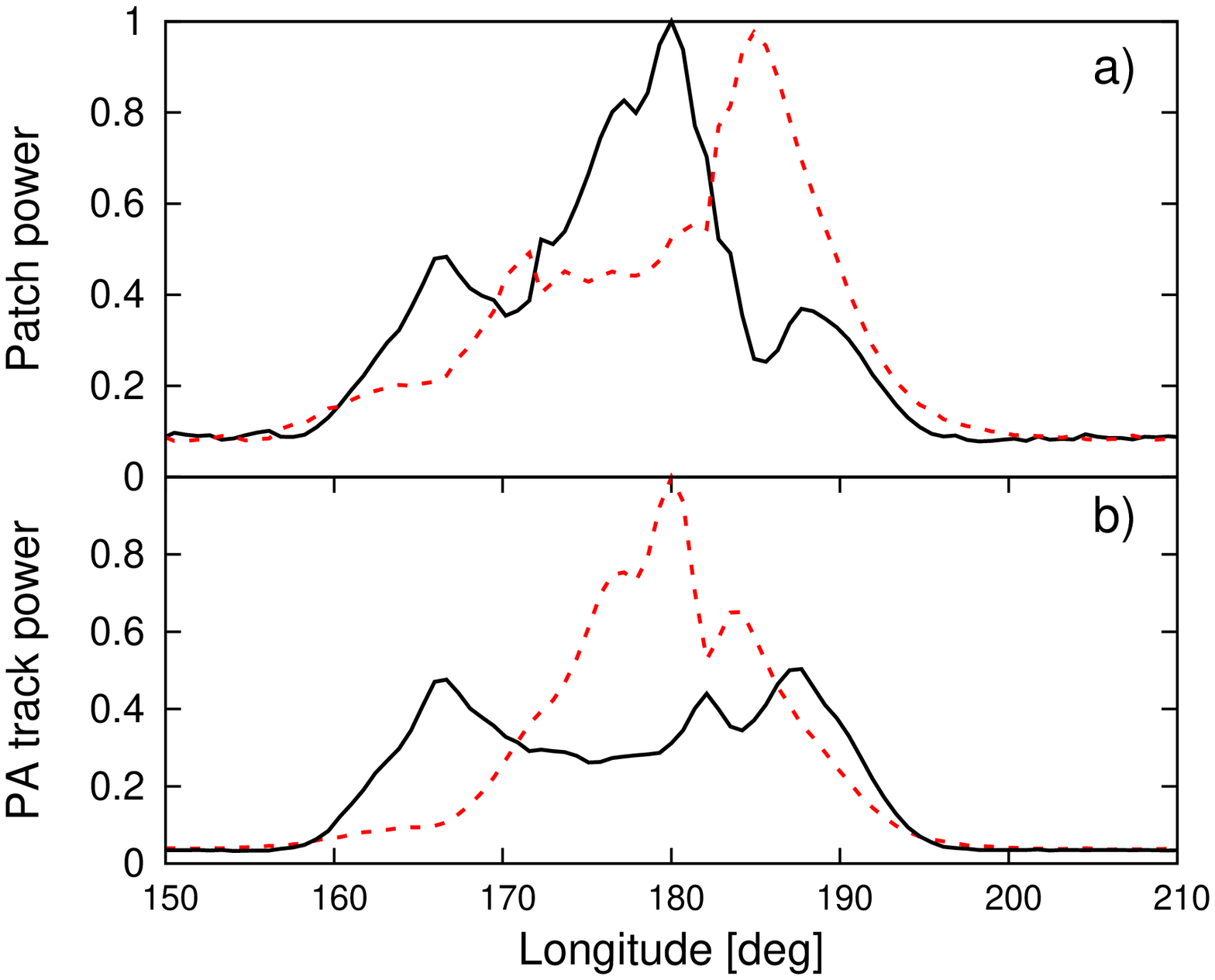}
\caption{
Top: normalised polarized flux integrated within each patch on the
P.~sphere. The solid line corresponds to patch 1 (mode 1). 
Bottom: normalised polarized flux integrated within each RVM-like track of
PA. The solid line is for the top track in Fig.~\ref{data1}e.
}
\label{modsep}
\end{figure}

The behaviour of polarization in PSR B1451$-$68 becomes clear when the
data are plotted on the Poincar\'e sphere (P. sphere). 
Fig.~\ref{data2} illustrates distributions of the polarization state on the P. sphere for 15 longitudes 
(a different set of 6 longitudes is shown in Fig.~3.45 of Ilie 2019). \nct{i19} 
 Each distribution consists of samples recorded in different rotation
periods and it has the form of two approximately antipodal flux
patches on P.~sphere. Hereafter these are called `patches' whereas the term
`modes' is reserved for the linear proper modes in high magnetic field. 
The question of when and whether 
the patches correspond to the proper modes or represent
a mixed state will be resolved in Section \ref{scomt}. 
Figure \ref{data2} implies that the patches are rotating almost meridionally 
on the P.~sphere, with each 
flux patch passing near each
V pole (northern and southern, ie.~positive and negative V). This gives four near-pole transitions which
correspond to the four peaks at $\kappa\sim\pm45^\circ$ in the bow tie curve of
the ellipticity angle (panel f of Fig.~\ref{data1}).

The labeling of the patches on the P.~sphere corresponds to those in the bottom two panels in Fig.~\ref{data1}. 
The stronger (primary) patch (no.~1) forms the upper PA track at the leading side of
the profile, but with increasing pulse longitude this patch moves downwards on the P.~sphere. 
After passing near the southern V pole, the flux of patch 1 forms the bottom PA
track (interval B in Fig.~1). Therefore, the upper PA track is stronger in interval A, whereas the
lower track is stronger in interval B. In the middle of interval B, the patches pass
through the equator ($\kappa \sim 0$), then they continue near-meridionally 
polewards, and pass near the V poles while entering the interval C. 
Therefore, the flux patch with no.~1 forms the peripheric
parts of the upper track while the track's center corresponds to patch no.~2. 
Opposite arrangement holds for the lower PA track (see the patch numbers in
Fig.~1). 
In the bottom panel in Fig.~1, patch 1 corresponds to
the `down-up-down' branch of the bow tie, whereas patch 2 to the `up-down-up'
branch. 

The $\pm 45^\circ$ tips of the bow tie coincide in longitude with the breaks
in the PA tracks, which is consistent with the V pole passage (hereafter VPP). 
The transition between interval A and B seems to exhibit an X-shaped
feature: the upper track PA decreases towards the lower track, whereas
the PA of the lower track increases towards the upper track, so that the 
tracks cross each other at the transition.
 Another transition -- between intervals B and C --  looks as a pale
break in the tracks, with the PA of a given track splitting both upwards and
downwards. This is most clearly seen on the left side of interval C for the
lower track.


The lower two panels of Fig.~\ref{data1} 
allow us to follow the strength (frequency) of each 
patch separately. The branches A1 and B1 are consistently stronger than A2
and B2 which is consistent with the V pole passage and the transfer of flux
between orthogonal PA tracks (see Fig.~2 in Dyks 2020).

On the trailing side of the profile, however, at the B-C transition, 
branch C2 becomes stronger than C1, ie.~patch 2
becomes stronger than patch 1. The coincidence of the bow tie tips at the B-C
transition with the breaks in the 
PA tracks leaves little doubts that the near-pole passage has occured at
$\Phi \approx 183^\circ$. Therefore, the radiative power of patch 1 must have been
transferred from the lower to the upper PA track at the B-C transition. 
The large strength of the C2 branch can thus only be explained by the
simultaneous inversion of the patch flux ratio, ie.~patch 2 becomes stronger 
than patch 1 on the trailing side of the profile (within interval
C).

It is therefore concluded that the inversion of power in the different parts
of PA track can occur in two different ways: either through the V pole
passage (whereby the power ratio in PA tracks is inversed, but not the ratio
of the patch flux content) or through the actual (real) change of power content
ie.~in each patch on the P.~sphere. 
In PSR B1451$-$68 the patch power inversion appears to coincide in longitude with
the V pole passage at the B-C transition. This is meaningful,
because the equal power (amplitude) of near-orthogonal waves is necessary for their
combination to produce the pure V polarization state.

The described behaviour is further complicated by the fact that the 
patches tend to assume an extended arc form near the V poles. Therefore,
while passing through the V pole, a single elongated patch can persist 
for a long time on opposite half-meridians, ie.~near the OPM transitions, 
a single patch can contribute to both orthogonal PA tracks `simultaneously'
(ie.~at a fixed longitude). 

Moreover, near the profile center (interval B) the modal patches assume
a double or bifurcated
shape, thus forming four subpatches. 
In Fig.~\ref{data2} the double form is clearly visible for the weak
(right) patch 2 at $\Phi=176^\circ$ and for the bright (left) patch 1 
at $\Phi = 180.2^\circ$. Each subpatch moves at a different speed in
different directions on the P.~sphere in such way that the subpatches 
coincide when crossing the equator, ie.~near the central
profile minimum at $\Phi=178^\circ$. Because of this
longitude-dependent bifurcation, each PA track in the center of Fig.~\ref{data1}e
(interval B) is split into two subtracks with different PA slope.  

 The shape-related complexities (extensions or bifurcations of patches) 
are addressed in Section \ref{deconf}.

\subsection{Profiles of separate polarization modes}

The power content of a given polarization mode can be defined
as the flux belonging to a given patch instead of the flux contained in a given RVM-like track
of PA. 
Fig.~\ref{modsep} presents mode-separated profiles calculated according to these two definitions.
The solid and dashed lines in the top panel present 
the polarized flux of patch 1 and 2 respectively. 
These are determined by integrating the polarized flux within each of the hemispheres 
of the P.~sphere after placing the centre of the brightest patch at the pole. 
One can see that most of the polarized power in the trailing side 
of the pulsar's profile is associated with patch 2. It is also evident in Fig.~\ref{modsep}a 
(as well in~Fig.~\ref{data2}) that
patch 2 also becomes slightly stronger during the first V passage.
Overall, the polarized profile of patch
2 (dashed) looks like a time-delayed version of the patch 1 profile (solid),
although there is no third trailing component in the former. 
Interestingly, two peaks in the flux profiles of different 
patches have nearly identical height. 
The patch-based mode separation produces profiles with a characteristic anticorrelation,
which is also observed in the modulation patterns of pulsars (Deshpande and
Rankin 2001; Ilie et al.~2020). \nct{iwj19, dr2001} 

The bottom panel of Fig.~\ref{modsep} presents the polarized flux integrated within 
each PA track, ie.~within $\pm45^\circ$ from an RVM fit. Such mode separation is similar
to the three-way mode segregation of Deshpande \& Rankin (2001). The method
produces completely different mode-segregated profiles than the patch-based
method: the top PA track (solid line) has a double form which does not dominate
 in the profie center. \nct{dr2001} 

The question on which panel of
Fig.~\ref{modsep} represents the correct mode separation depends on the
physical origin of the VPP effect. If the rotating patches
correspond to elliptic proper orthogonal polarization modes, then Fig.~\ref{modsep}a is
appropriate. 
If, however, the rotating patches only represent a mixed state that
results from the coherent superposition of linear proper modes, then
Fig.~\ref{modsep}b applies. In the latter case the VPP just represents a coherent OPM
jump, ie.~the proper polarization states do not pass near the V poles.
 In Section \ref{scomt} we show that the coherent OPM jump is the correct interpretation.
This implies that adjacent profile components have to consist of linear 
orthogonal modes with appropriate phase lag difference (see Section \ref{overlap}).

\subsection{Polarization fractions and similarity to PSR B1237$+$25}

The strengths of the two tracks (and patches) are comparable for PSR B1451$-$68, 
therefore, the polarization fractions
calculated by averaging of Stokes parameters are low, and much lower
than the instantaneous polarization fractions of individual samples. Panel b of
Fig.~\ref{data1} presents arithmetic average of instantaneous polarization
fractions, ie.
\begin{equation}
L/I \equiv N^{-1}\sum_{i=1}^N L_i/I_i = N^{-1}\sum_{i=1}^N
\sqrt{Q_i^2+U_i^2}/I_i 
\end{equation}
\begin{equation}
|V|/I \equiv N^{-1}\sum_{i=1}^N |V_i|/I_i  
\end{equation}
\begin{equation}
I_\mathrm{pol}/I \equiv N^{-1}\sum_{i=1}^N \sqrt{Q_i^2+U_i^2+V_i^2}/I_i
\end{equation}
where $N$ is the number of samples at a longitude $\Phi$ and the index
$i$ refers to a given rotation period. The fractions are shown with the red,
green and blue dashed lines in Fig.~\ref{data1}b. These averages are biased quantities in the presence of noise. The values of $L/I$ and $I_\mathrm{pol}/I$
were debiased following the correction of Wardle \& Kronberg (1974). \nct{wk74}
The effect of the bias is further reduced by discarding the weakest samples, but a bias remains. Nevertheless, there are structures apparent which indicate relative changes in the arithmetic averages as function of pulse longitude.
 Unlike these `typical instantaneous' quantities, 
the polarized fluxes shown in Fig.~\ref{data1}a, have been calculated in the
usual way, ie.~by averaging the Stokes parameters.

Fig.~\ref{data1}b shows that near the V poles ($\Phi=172^\circ$ and
$183^\circ$) the instantaneous polarization degree is typically quite high
($I_{\rm pol}/I\approx0.5 - 0.6$). Therefore, the incoherent mode
superposition (Melrose et al.~2006) is not responsible for the observed VPP
(though it is at work in other objects).

The `typical instantaneous' polarization fractions reveal
a quite symmetric form, with twin minima in $L/I$ flanking the maximum of the core
component. The minima coincide with peaks of $|V|/I$ which also has a quite
symmetric form. These curves reflect the near-meridional rotation of
polarization state, which produces the anticorellation between $L/I$ and
$|V|/I$. 

Thus, regarding the typical instantaneous polarization fractions, the pulsar
B1451$-$68 resembles the well known `complex core' pulsar B1237$+$25, for which
the V-coincident twin minima have also been observed and interpreted through
the rotation of the polarization state (D17, D20). In B1237$+$25, however,
the features are visible in the standard Stokes-averaged fractions, probably
because of larger disproportion of OPMs' power. In the case of
B1451$-$68, the standard Stokes-averaged polarized fractions have complex
longitude dependence and do not exhibit any clear symmetry (so we do not show
them in this paper).

\section{Analytical formulae for V pole passage} 

 In this section we introduce the equations for the rotation of a pol.~state on
Poincar\'e sphere. It is shown that the rotation implies essentially the
same PA curve as the RVM. In the following Section \ref{scomt}, it is shown
that the rotation is caused by the coherent orthogonal mode transition,
ie.~by variations of the ratio of mode amplitudes. Coherent OPM jumps do 
not necessarily involve the decrease of the linear polarized fraction.
Therefore, several OPM jumps have been misinterpreted as RVM and many of them 
might have been missed in pulsar data (Section
\ref{scomt}).

\begin{figure}
\includegraphics[width=0.48\textwidth]{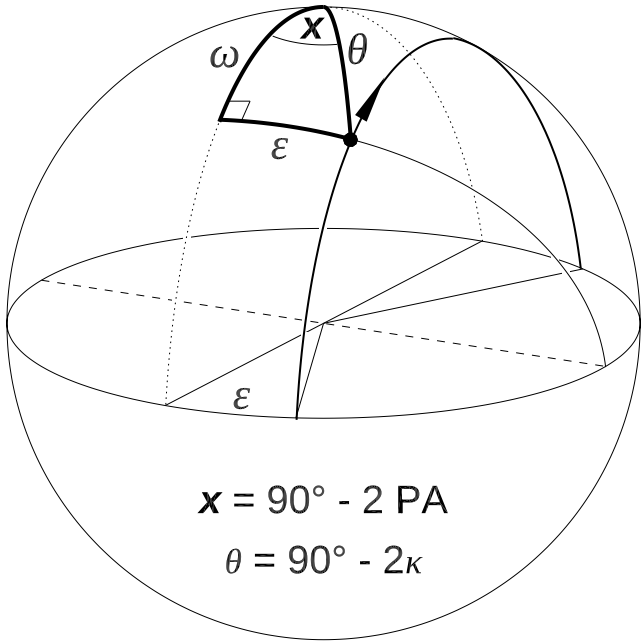}
\caption{Geometry of the near meridional circularization on the Poincar\'e
sphere in the case of linear retarder axis (equatorial axis of patch rotation, dashed diameter). 
The polarization state (flux patch on the Poincar\'e sphere) moves along the
thick circle parallel to the dotted meridian. The position of the state is
parametrised by the angle $\omega$. 
The thick spherical triangle 
is used to derive eqs.~(\ref{paone}) and (\ref{elione}). 
}
\label{pass}
\end{figure}

The rotation of the polarization state on the Poincar\'e sphere may reflect
intrinsic properties of the emission or it may directly result from a coherent
superposition of orthogonally polarized waves that oscillate at some relative phase lag. 
Regardless of the actual origin, the patch rotation can be mathematically considered as a retarder action
on the Poincar\'e sphere, with the rotation angle corresponding to the phase
lag. The purely meridional patch motion (meridional circularization)
 may in particular correspond to a linear retarder action, with the polarization state
rotating around the axis contained in the equatorial plane ($Q$-$U$ plane).
 Physically this may correspond to the O-mode retardation (Edwards
\& Stappers 2004; Jones 2016; D17). \nct{es04, jon2016, d2017} 
Aside from this mechanism, the meridional circularization may also correspond to
a coherent OPM transition
(inversion of mode amplitude ratio at a fixed quarter-wave phase lag). 

\begin{figure}
\includegraphics[width=0.48\textwidth]{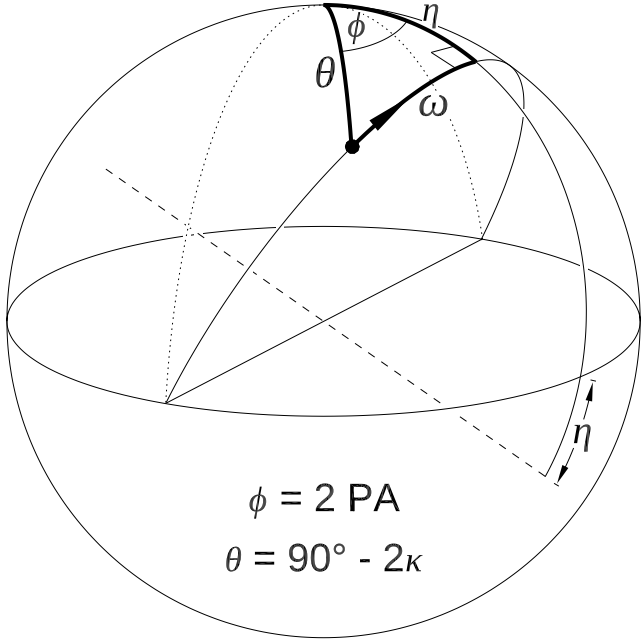}
\caption{Geometry of near meridional circularization in the case of
an arbitrary axis of patch rotation (dashed diameter). The polarization state
(bullet) moves along the thin solid great circle and the states' position is parametrized by the
angle $\omega$. 
The thick spherical triangle 
is used to derive eqs.~(\ref{patwo}) and (\ref{elitwo}). 
}
\label{passtwo}
\end{figure}

Before introducing a general case, this section deals with two simple limiting cases of pole
passage:  
1) Rotation of arbitrary polarization state around an equatorial axis (linear
retarder, see Fig.~\ref{pass}). Here the arbitrariness of polarization state implies that motion along small
circles is allowed, ie.~the angle between the polarization state and
the rotation axis may be arbitrary. 2) Great-circle rotation around a nonequatorial axis
(coherent OPM jump, ie.~transition between two antipodal modes; also a retarder with elliptical proper
modes). In this case only the motion along a great
circle is considered (see Fig.~\ref{passtwo}). Finally, the two cases are combined in a general case of 
rotation in arbitrary circle (ie.~small or great) around an arbitrary rotation axis.

\subsection{Linear retarder, arbitrary polarization state}

In this case the polarization state (flux patch represented by the bullet in
Fig.~\ref{pass}) is being rotated by an angle $\omega$ around a rotation axis that is contained within
the equator of the Poincar{\'e} sphere (dashed diameter). 
The patch is passing at an angular distance $\epsilon$ from the V
pole.\footnote{The value of $\epsilon$ does not have to be small for the
derived equations to
hold.} 
The polarization angle
$\psi$ is equal to half the azimuth, hence it is related to the angle
$x$ through $x=90^\circ-2\psi$. The colatitude $\theta$ is related to the
ellipticity angle $\kappa$ through $\theta = 90^\circ -
2\kappa$. Using these relations and the standard formulae of spherical
trigonometry, it is easy to find the equation for the polarization angle: 
\begin{equation}
\tan(2\psi) = \frac{\sin\omega}{\tan\epsilon}
\label{paone}
\end{equation}
and the ellipticity:
\begin{equation}
\sin(2\kappa) = \cos\omega \cos\epsilon.
\label{elione}
\end{equation}
Here the rotation angle is some unknown function of time. In the simplest case
$\omega = C(\Phi-\Phi_1)$, where $\Phi$ is the pulse longitude and $C$ and $\Phi_1$
are free parameters. This is equivalent to considering $\omega$ as the pulse longitude
in arbitrary units. $\Phi_1$ is the pulse longitude at the middle of the VPP OPM
transition ($\omega=0$ ie.~the closest approach to the V pole). 
Note that eq.~(\ref{paone}) represents the value of PA
which is measured with respect to the PA of the patch rotation axis
$\psi_\mathrm{ra}$, ie.~$\psi_\mathrm{obs} = \psi +
\psi_\mathrm{ra}$. In simple empirical models of coherent mode superposition the patch
rotation axis is defined by proper modes that correspond to the magnetic field in the polarization limiting region 
(D17). Therefore, $\psi_\mathrm{ra}$ = $\psi_\mathrm{RVM} + \psi_0$, where $\psi_0$ is a 
constant free parameter.
In the case when two quasi-orthogonal RVM tracks are observed at a PA of $\psi_1$
and $\psi_2$, the symmetry of Fig.~\ref{pass} implies that $\psi_\mathrm{ra}=
(\psi_1+\psi_2)/2$, ie.~the PA of patch rotation axis is half way between
the observed PA tracks. This is because the PA values in the two RVM tracks correspond to 
two bullet positions in the equatorial plane.  

Eq.~(\ref{paone}) describes the passage of radiative power from one PA track to the other,
near-orthogonal PA
track. Thus, eq.~(\ref{paone}) is followed during an OPM jump when the jump is
caused by the V pole passage.  
This equation has the geometric form of an S-shaped function, therefore an
observed OPM jump caused by the V pole passage can be
mistaken for the RVM S-swing.


\subsection{Coherent OPM jump at a fixed phase lag; elliptical retarder with great circle rotation}

In the case of elliptical proper modes the patch is rotated around 
an arbitrarily oriented axis (dashed diameter in Fig.~\ref{passtwo}). Here the
axis is tilted by an arbitrary angle $\eta$ with respect to the V axis.
The polarization angle $\psi = \phi/2$, where $\phi$ is
the azimuth (Fig.~\ref{passtwo}).  Again, the value of
$\psi$ is the PA value as measured with respect to the PA of the patch rotation
axis,
ie.~$\psi_\mathrm{obs} = \psi + \psi_\mathrm{RVM}+\psi_0$.

 In the considered case the thick-line spherical triangle in Fig.~\ref{passtwo} implies
\begin{equation}
\tan(2\psi) = \frac{\tan\omega}{\sin\eta},
\label{patwo}
\end{equation}
whereas the ellipticity angle has the same form as before: 
\begin{equation}
\sin(2\kappa)=\cos\omega\cos\eta.
\label{elitwo}
\end{equation}

 As explained in Section \ref{scomt}, the rotation geometry of
Fig.~\ref{passtwo},
ie.~the great circle rotation between antipodal points, also refers
to the mode-ratio driven coherent OPM transition (COMT), however, 
in the COMT case the proper mode axis corresponds to the solid diameter in
Fig.~\ref{passtwo} and the modes are linear, not elliptic. 
The mode strength ratio and phase lag $\dox$ correspond to
different angles in Fig.~\ref{passtwo} than in the case of retardation (see
Fig.~\ref{comt} and Sect.~\ref{scomt}).

\begin{figure}
\includegraphics[width=0.48\textwidth]{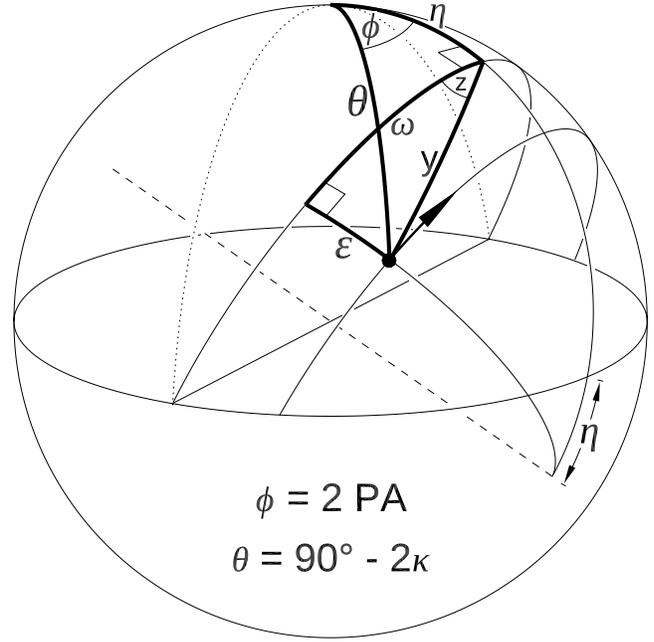}
\caption{Geometry of circularization in the case of
arbitrary polarization state (bullet) and arbitrary axis of the state's rotation (dashed diameter). 
The derivation of eqs.~(\ref{pagen}) and (\ref{eligen}) makes use of the two thick spherical 
triangles: $(\theta, \eta, y)$ and $(\epsilon, \omega, y)$, where the shared
side $y$ cancells out in the process.  
}
\label{passtri}
\end{figure}

\subsection{General case: arbitrary-circle rotation around arbitrary axis} 

 To obtain a general result,  
the rotation axis (dashed line in Fig.~\ref{pass}) needs to be inclined from the equatorial plane by 
the angle $\eta$ (Fig.~\ref{passtri}). 
The azimuth $\phi$ and colatitude $\theta$ must then be
redefined to refer to the new dislocated bullet position. 
With the standard 
spherical trigonometry procedures one may then find the following ellipticity
angle:
\begin{equation}
\sin(2\kappa) = \cos\theta = \cos\eta\cos\omega\cos\epsilon - 
\sin\eta\sin\epsilon
\label{eligen}
\end{equation}
and the following PA:
\begin{equation}
\tan(2\psi) = \frac{\cos\epsilon\sin\omega}{\cos\epsilon\cos\omega\sin\eta +
\sin\epsilon\cos\eta}.
\label{pagen}
\end{equation}
As usual 
\begin{equation}
\psi_\mathrm{obs} = \psi + \psi_\mathrm{RVM} + \psi_0.
\label{pasum}
\end{equation}
Eq.~(\ref{pagen}) is a general formula which describes an OPM jump 
 which in particular may involve the V pole
passage. 
In the limit of $\eta=0$ or $\epsilon=0$, eq.~(\ref{pagen}) reduces to
eqs.~(\ref{paone}) or (\ref{patwo}), respectively.

It must be emphasized that for a linear change of $\omega$ with pulse
longitude $\Phi$, the functional dependence of eq.~(\ref{pagen}) is
essentially the same as the RVM equation:
\begin{equation}
\tan(\psi_\mathrm{RVM}) = \frac{\sin\alpha\sin(\Phi-\Phi_0)}{\cos(\Phi-\Phi_0)\cos\zeta\sin\alpha -
\cos\alpha\sin\zeta}.
\label{rvm}
\end{equation}
For the following substitutions: $\alpha = 90^\circ+\epsilon$, $\zeta
=90^\circ-\eta$, the azimuth on the Poincar\'e sphere  (twice the $\psi$ value)
changes in exactly the same way as the PA predicted by the RVM.  

As before, some observed S-shaped variations of PA may be mainly caused by the VPP
rather than by the RVM. This may happen when the VPP variations of PA are much
faster than RVM, ie.~the RVM curve is relatively flat. 
In such case the S swing should be modelled with eq.~(\ref{pagen}). 
When the RVM-driven changes of PA are also noticeable, eq.~(\ref{pasum}) needs to be used.

\section{Coherent orthogonal polarization mode transition} 
\label{scomt}

\subsection{Coherent OPM jump or retardation?}

The main possible interpretations of the meridional circularization involve
the retardation of the O mode and the coherent OPM transition. 
In the case of linear proper modes, the increase of the phase lag $\dox$
between the X and O mode causes 
the rotation of the summed polarization state around the equatorial QU axis
(Fig.~\ref{pass}). The lag is changing along with the $\omega$ angle: $\dox=90^\circ-\omega$. 
If the amplitudes of the modes $E_X$ and
$E_O$ are equal, the bullet follows the meridian which is orthogonal to
the dashed proper mode axis ($\epsilon=0$). Otherwise, $\epsilon=90^\circ-2\pmx$, where
$\tan\pmx=E_O/E_X$ is the mixing angle. 
A spread of phase lags results in a spread of polarization state vectors in
the meridional direction, which seems
consistent with the patch extention along the meridian, as observed near the
V poles. However, this is the only feature that may be considered consistent
with the observations, whereas multiple arguments can be put against the retardation. 

First, the meridional retardation requires that the mode strength
stays equal throughout the entire phenomenon, whereas the mode ratio is
typically variable in pulsar signal (both in pulse longitude and in modulation
phase).

Second, when the patches are far from the QU equator, we never observe the proper polarization states separately. 
If observed alone, the modes would produce two additional flux patches in directions orthogonal to the bullet
motion plane (ie.~at the tips of the dashed diagonal in Fig.~\ref{pass}).

Third, in the course of the increasingly strong retardation, the observed polarization state
(bullet in Fig.~\ref{pass}) is always a
mixed state: it is produced by the coherent superposition of the proper modes O and
X. The bullet never coincides with the X or O mode. Contrary to this fact,
the patches observed in B1451$-$68 become very compact while crossing the QU equator,
which can be understood as a coincidence of the observed patches with the linear proper modes 
(transition from a mixed to the pure polarization state -- see below). 

Fourth, if the phase lag $\dox$ is acquired through propagation in 
birefringent medium, then, given the tubular symmetry of the polar region,
it is natural to expect that $\dox$ is a nonmonotonic function of $\Phi$, 
eg.~$\dox$ could reach a maximum in the middle of the profile and vanish at
the profile edges.  
In contrast to this expectation, the bow tie form of the polarization in B1451$-$68 clearly
implies that the retarder rotation angle is steadily increasing across the full
pulse window (from, say, $\dox=0$, through $180^\circ$, up to almost 
$360^\circ$). The modal patches never stop rotating (as expected when the phase lag reaches
maximum) nor they move backward, as expected when the birefringent medium
becomes rarefied towards the edges of the profile.  Such lack of
mirror symmetry could possibly result from ray propagation at high altitudes, but 
the simplest model based on the retardation cannot explain these observations. 


It is therefore concluded that the meridional circularization is caused by
the coherent OPM transition.

\begin{figure}
\includegraphics[width=0.48\textwidth]{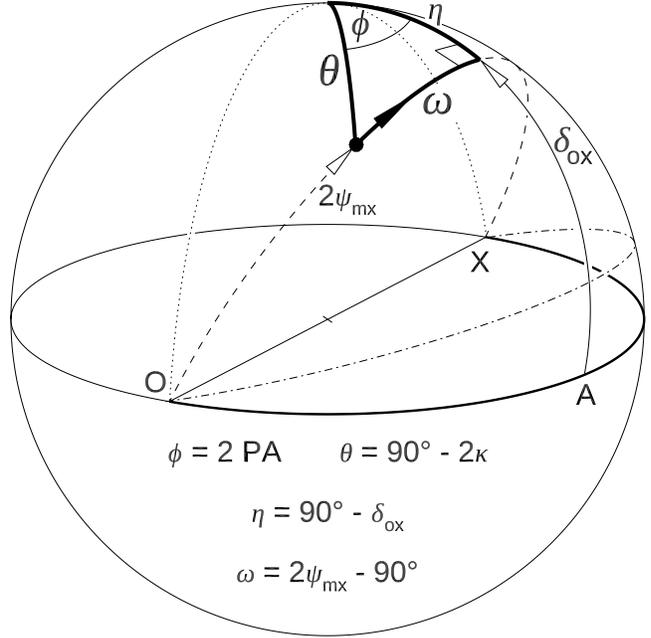}
\caption{Geometry of the coherent OPM transition on the P.~sphere. The
change of mode ratio corresponds to the change of mixing angle $\pmx$. 
For a constant phase lag $\dox$ the polarization state (bullet) follows a
great circle tilted at the angle $\dox$ with respect to the QU equator. The
state can move meridionally through the V pole ($\dox=90^\circ+n180^\circ$) or
equatorially through the A point ($\dox=n180^\circ$).
}
\label{comt}
\end{figure}

\subsection{Coherent OPM transition} 
 
The geometry of the coherent OPM transition on the P.~sphere is shown in
Fig.~\ref{comt}. The bullet corresponds to the coherent sum of the linear X and O
modes of varying amplitude ratio. The mode amplitudes can be written as
$E_X=E\cos\pmx$ and $E_O=E\sin\pmx$ so that
the ratio of the mode amplitudes is parametrized with
the mixing angle $\tan\pmx=E_O/E_X$. Definitions of the Stokes parameters
lead to eqs.~(4)-(7) in D17 for the coherent sum of the modal waves. 
After substitution of the amplitudes (and change of the V sign convention as
compared to that used in D17\footnote{The sign of V was chosen so that
$\dox$ as marked in Fig.~\ref{comt} is positive. The angle
$2\pmx=\omega+90^\circ$ is defined so that the value of $\omega$ shown in
Fig.~\ref{comt} is negative.})  
the Stokes parameters for the coherent sum become: 
\begin{eqnarray}
I&=&E^2\\
Q/I&=&\cos(2\pmx)\label{qeq}\\
U/I&=&\sin(2\pmx)\cos\dox\\
V/I&=&\sin(2\pmx)\sin\dox.\label{veq}
\end{eqnarray}
Eqs.~(\ref{qeq})-(\ref{veq}) imply that $\pmx$ is half of the angle measured from
the OX axis along a great circle followed by the bullet (dashed path in
Fig.~\ref{comt}), ie.~$\omega=2\pmx-90^\circ$. The inclination of the great
circle is fixed by the value of the OX phase lag $\dox=90^\circ-\eta$.

By substituting for $\omega(\pmx)$ and $\eta(\dox)$ in eqs.~(\ref{patwo}) and (\ref{elitwo}) 
the following equations for the coherent OPM jump are obtained: 
\begin{equation}
\tan{(2\psi)}=\frac{\tan{(2\pmx-90^\circ)}}{\cos\dox},
\label{pacomt}
\end{equation}
\begin{equation}
\sin{(2\kappa)}=\sin{(2\pmx)}\sin\dox.
\end{equation}
Eq.~(\ref{pacomt}) has nearly the same form as the RVM for $\alpha=90^\circ$ and
$\zeta=\dox$:
\begin{equation}
\tan{\psi_{RVM}}= \frac{\tan{(\Phi-\Phi_0)}}{\cos\zeta}.
\end{equation}

If $\dox\sim90^\circ+n180^\circ$ the bullet follows the meridional
path (dotted vertical in Fig.~\ref{comt}). This involves the V pole passage at
$\pmx=45^\circ$ associated with the OPM transition between the PA tracks. 
This type of quarter-wave OPM jump can be recognized by the very high
$|V|/I\gg L/I$ coincident with minima in $L/I$ 
(both observed at the longitude of the OPM jump).

During the meridional motion, the PA stays constant until the pole is passed by. 
Therefore, the corresponding PA curves are flat
between the VPP OPM jumps. In case of two OPM jumps (eg.~on both sides of
a core component), the PA tracks can have the shape of stairs (two upward OPM jumps, or
two downward), the U
shape (downward and upward OPM jump) or a reversed U shape (jump up and
down). Given that this quarter-wave OPM transition, as any coherent OPM
transition, is described by the RVM-like equation, the jump can sometimes be 
mistaken with the RVM, see the next subsection for examples.

If the lag $\dox$ does not precisely have the quarter-wave value, the pol.~state
can move along the dashed line in Fig.~\ref{comt}. 
This corresponds to a PA curve that has the shape of stairs with slanted treads. 
If $\dox$ is even further from the quarter-wave value (dot-dashed line in
Fig.~\ref{comt}, the stairs' treads become more slanted and the OPM jump is
more gradual. Also in this case the maxima of $|V|/I$
coincide with minima in $L/I$, however, in this type of coherent jump  $|V|/I < L/I$. 
This case is observed, eg.~for B1913+16 (Weisberg \& Taylor 2002; fig.~1 in Dyks 2017).
\nct{wt02, d2017} 

The most deceiving coherent OPM transition, which is most difficult to notice
in pulsar data, corresponds to no lag, or to half-wavelength lags ($\dox=n
180^\circ$). The pol.~state is then moving along the thick part of the QU equator
in Fig.~\ref{comt}. The polarized fractions do not change at all during this
OPM jump ($L/I=I_{\rm pol}/I=1$, $V/I=0$). Moreover, the PA is changing
gradually during such half-wave OPM jump, being always equal to the mixing angle, as
determined by the mode amplitude ratio. It is extremely difficult to
discern this type of a coherent OPM jump from a PA variation of another origin,
eg.~the RVM. If $\dox\ne n 180^\circ$ the OPM transition can be recognized
through the flattened stairs-shaped (or U-shaped) PA curves and through the
peaks of $|V|/I$ that coincide with minima of $L/I$. Both these signatures are unavailabe in the case of the
half-wave OPM jump ($\dox=n 180^\circ$). This is the type of OPM jump, to which we are completely blind.

\subsection{Misinterpretations of S-shaped PA swings}

\subsubsection{Quarter-wave OPM transitions}

The observed S-shaped swings may thus have two origins: they are either RVM driven, 
or result from a coherent OPM jump (COMT).  
Because RVM and COMT have nearly the same functional form, some 
observed S swings have been described in the literature as RVM 
instead of COMT. 
 Even in the case of the quarter-wave OPM jump, such misinterpretation is likely, because the
swings of RVM origin are identified on longitude-PA diagrams through 
the continuity of their PA track. Since the VPP OPM involves a single
flux patch on Poincar\'e sphere, a similar power on both sides of the OPM jump is ensured in the patch's PA 
track,\footnote{As discussed in D20 \nct{d2020} (and further below),
the passage through the V pole spreads the radiative power over all PA values
which makes the quarter-wave OPM transition weak on longitude-PA diagrams. However, when
most power is passing on one side of the V pole 
($\dox\sim90^\circ+n180^\circ$), the PA track stays strong
through the OPM transition.} which can easily make the impression of an RVM S swing. 
Moreover, the fixed sign of V at the transition furthermore supports
the erroneous impression that the observed S swing is not an OPM jump.

Since the S swings of the quarter-wave COMT origin are orthogonal transitions between
the PA tracks, they should span about $90^\circ$ in PA. This, however, may be
insufficient to recognize them on longitude-PA plots, because the VPP OPM transitions 
coexist with the simultaneous RVM effect, so the actual difference of PA on both sides of
the S swing's center may be different from $90^\circ$.\footnote{In the case
of the retardation-driven OPM jumps (Fig.~\ref{pass}), the further from the V
pole the passage is, the less orthogonal is the transition, even in the
absence of RVM.}

 In the case of the quarter-wave OPM jumps, they can be recognized by the 
high degree of circular polarization $|V|/I$. In
the average profiles, however, the presence of 
the antipodal patch on the P. sphere can effectively decrease the
average $|V|/I$ to very low values (as in B1451$-$68, top panel in
Fig.~\ref{data1}). Moreover, $V/I$ can be suppressed by the spread of
polarization states (which can be caused by the mode ratio spread or phase lag
spread). 
Therefore, when viewing pulsar data, 
it is useful to inspect and illustrate the arithmetic average
of the instantaneous polarization degree as a function of pulse longitude 
(Fig.~\ref{data1}b),  
 in addition to the usual Stokes-averaged polarization degree. 

The additional plot of ellipticity angle (Ilie 2019) can help resolving the issue, but
the most certain way to discern the swing's
nature (RVM vs the quarter-wave COMT) is to study the polarization on the Poincar\'e sphere
(or better yet in the Stokes space). If the polarization of the individual pulses is undetectable,
other techniques need to be used to probe distributions of polarization
states in the Stokes space (van Straten \& Tiburzi 2017). 
\nct{vst17}

 The cases where the VPP OPM transitions are modelled as an RVM swing can be found
in pulsar literature and include, eg.~PSR J1841$-$0500 ($\Phi = -2^\circ$ in
fig.~3 of Camilo et al.~2012), PSR B1237$+$25 ($\Phi\approx-0.5^\circ$ in
fig.~1 of Smith et al.~2013), PSR B1857$-$26 (J1900$-$2600) and B1910$+$20 ($\Phi = 1^\circ$ in the
extended online version of fig.~A7 in
Mitra \& Rankin 2011). In all these cases the polarization degree is not
negligible and $|V|/I \gg L/I$. 
\nct{srm13, mr2011, crc2012}

\subsubsection{Similarity to B1857$-$26}

In the profile of PSR B1857$-$26 (J1900$-$2600) there are two COMT-related 
transitions on both sides of the core component (at $\Phi=-4^\circ$ and
$\Phi=3^\circ$ in fig.~2 of Mitra \& Rankin 2008). Within the core the
strong PA track is only roughly following an RVM curve orthogonal to the PA
track observed in both peripheries. On the leading side
the polarization state is passing at a somewhat further distance from the V
pole: at
$325$ MHz this looks like a $45^\circ$ PA jump. The trailing VPP is closer to
the V pole (fig. 3.64 in Ilie 2019), so the trailing PA jump is closer to
the orthogonal one. 
The core part of the observed PA track is strongly affected by the frequency-dependent
 coherent OPM superposition  (the PA slope is $\nu$-dependent, see fig.~6 in Johnston et
al.~2008). \nct{jkm2008}
Since the leading-side PA track follows different RVM 
than in the center, neither RVM fit in Mitra \& Rankin
(2008) is justified. Although their solid line fit is correctly passing from the
`primary' to the `secondary' PA track on the core's trailing side, 
both fits ignore the presence of the coherent mode transition on the leading-side.
The evolution of this profile (J1900$-$2600) with frequency
$\nu$ is shown in the left column 
of fig.~6 in Johnston et al.~(2008). \nct{jkm2008} The profile undergoes  
 transformation from the B1451$-$68-like form at high $\nu$ to the B1237-like form at low $\nu$. 
At high $\nu$ the profile is boxy and the sinusoidal V profile 
is almost as wide as the Stokes I profile, similar to PSR B1451$-$68, except that 
one patch dominates in J1900$-$2600. 
Along with decreasing $\nu$, clear minima in total intensity appear on both sides of the core,
and the central sign-changing of V is squeezed to a narrow longitude
interval under the core.  Whatever is their physical origin, 
the sinusoid-like V profiles within cores are related to the profile-wide
sinusoidal V profiles as observed in different objects or in the same object at a higher
frequency.

\begin{figure}
\includegraphics[width=0.48\textwidth]{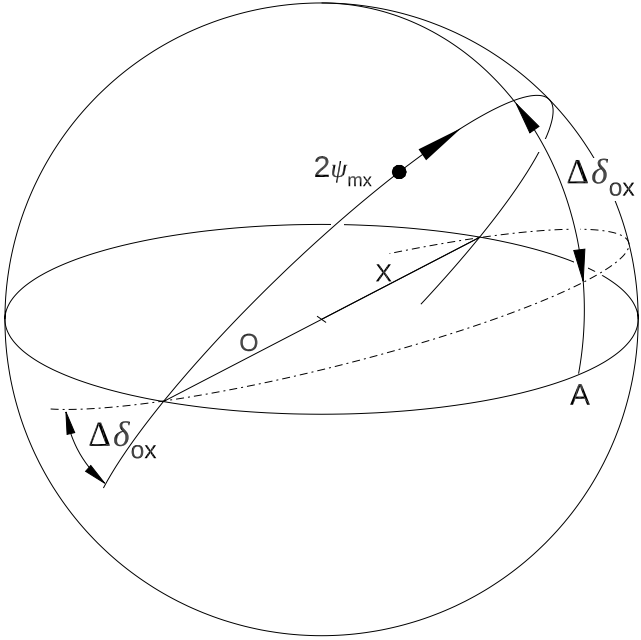}
\caption{The cause of the observed patch compactness in the QU equator. 
A spread in the phase lag $\Delta\dox$ extends the patches beyond the
equator and makes them compact near the OX proper mode axis. 
}
\label{bif}
\end{figure}

\subsubsection{Half-wave OPM transitions}

It must be emphasized that the above examples of the quarter-wave coherent OPM jumps 
were selected because they are easy to identify. 
It is also possible to identify intermediate cases of OPM jumps, with
$L/I>|V|/I$, which correspond to the dot-dashed path in Fig.~\ref{comt}. 
These can be recognized by the anticorellation 
of $L/I$ and $|V|/I$: maxima of $|V|/I$, which is low everywhere in the
average profile, coincide with minor minima in $L/I$, which is high everewhere
in the profile.  
This intermediate jump type can also be recognized through the PA curve with the shape
of stairs with slanted treads. B1913+16 (Weisberg \& Taylor 2002; fig.~1 in Dyks 2017) 
\nct{wt02, d2017} 
is an example of such intermediate-lag COMT. 

However, neither the PA curve nor the polarized fractions reveal 
the half-wave OPM jumps ($\dox\sim n180^\circ$). Since 
it was easy to find 
the previous types of
coherent OPM transitions (quarter-wave and intermediate) in pulsar data, 
it must be concluded that the half-wave transitions are
likely present in pulsar signals too. Therefore, radio pulsar
data can include numerous unrecognized coherent OPM jumps. 
Since the half-wave coherent OPM transitions stay highly linearly polarized
and only involve the gradual change of PA, it is likely that
several S-swings are actually the half-wave COMTs. 

It is shown in accompanying paper that this finding has far-reaching
consequences,
because it allows for previously forbidden interpretations of pulsar
polarization: the OPM transitions can be claimed for longitudes
and frequencies without vanishing polarized fractions.

\subsection{Patch deconfinement beyond the QU equator.}
\label{deconf}

As described in Section \ref{patches} (Fig.~\ref{data2}), the observed patches become compact when
crossing the QU equator. The confinement can be understood as the result of 
finite spread of the OX lag, as explained in Fig.~\ref{bif}. 
For a specific value of $\dox$, the transition between the mode O and X (or vice versa)
occurs along a great circle with diameter along the proper mode
axis. The OPM transitions that correspond to different $\dox$ values follow
different great circles, all of which pass through the same proper mode
diameter, as shown in Fig.~\ref{bif}. Therefore, the patches are compact
when they coincide with the proper modes, whereas they  
become extended beyond the QU equator. More precisely, 
the patches extend far from the proper mode axis, therefore, for the
half-wave COMT ($\dox=n180^\circ$), they could become extended near point A. 
This seems to be the only way to discern between the half-wave COMT and RVM.

It is thus possible to answer the question on whether the rotating patches are proper modes or a mixed
state: they are either of these at different
pulse longitudes. If $\dox \ne n180^\circ$, the patches present linear proper modes only when
passing through the QU equator, whereas beyond the equator they represent a mixed state
(coherent superposition of the proper modes). 

\begin{figure}
\includegraphics[width=0.48\textwidth]{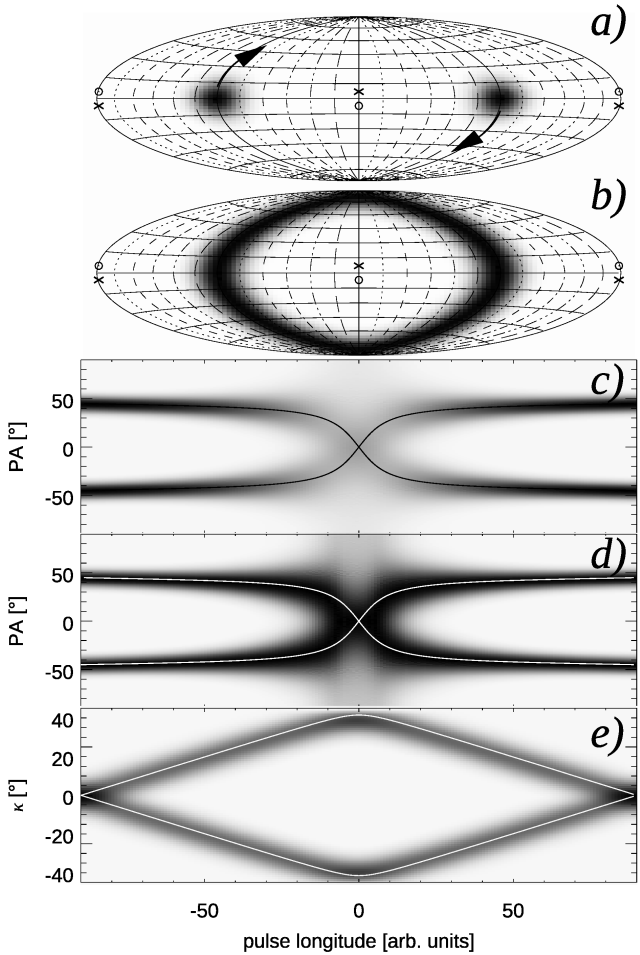}
\caption{An X-type OPM transition for two patches of equal strength
that pass near the V pole ($\eta=\pm7^\circ$).  
{\bf a)} Initially-orthogonal location of modal patches on P.~sphere 
(Gaussian patches with $1\sigma$ width of $10^\circ$, located at azimuths $\pm
90^\circ$, Hammer equal area projection). The patches
are rotated around two different axes that pierce the sphere at the
`o' marks (left patch) or `x'
marks (right patch). {\bf b)} Trace of the patches after their uniform rotation by
$180^\circ$. {\bf c)} The corresponding view of the OPM transition on the standard
longitude-PA diagram. Solid lines show the analytical model of
eq.~(\ref{pagen}). 
{\bf d)} Same as c), but the grey scale is normalized
for each longitude separately (black = maximum value, white = zero). {\bf e)}
The corresponding ellipticity angle. White lines follow
eq.~(\ref{eligen}). 
The quantity represented by the
grey scale may be considered as the histogram of the frequency of occurence,
or cumulative flux. 
}
\label{five}
\end{figure}

\begin{figure}
\includegraphics[width=0.48\textwidth]{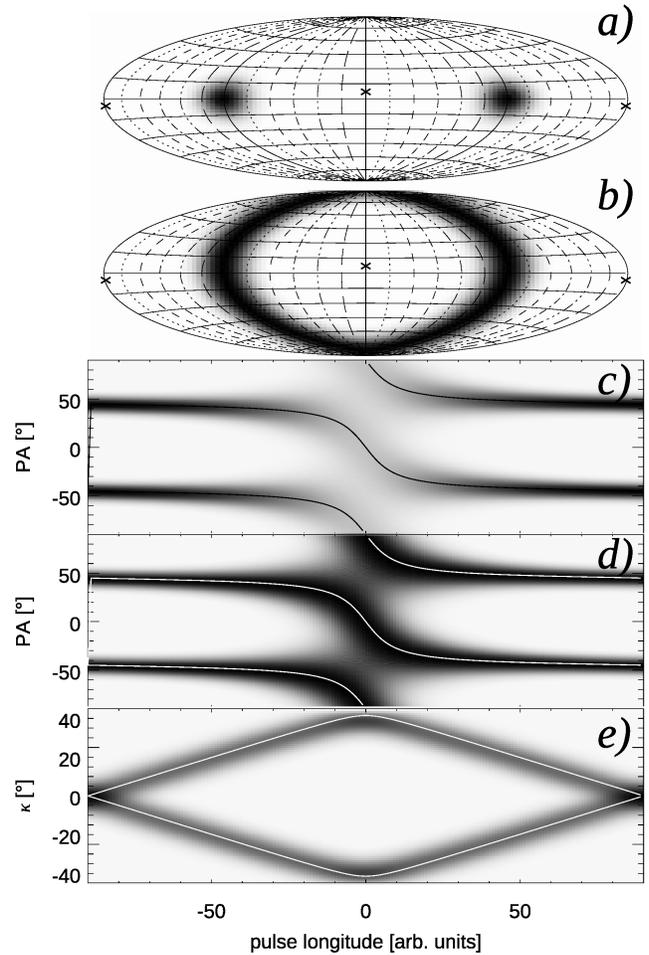}
\caption{Same as in Fig.~\ref{five} except now both patches 
rotate around the same rotation axis with $\eta=7^\circ$
(`x' marks). Therefore, the patches are passing on
the opposite sides of the V poles. This leads to the double S-shaped orthogonal
polarization mode transition (panels c and d).
}
\label{fivec}
\end{figure}
\begin{figure}
\includegraphics[width=0.48\textwidth]{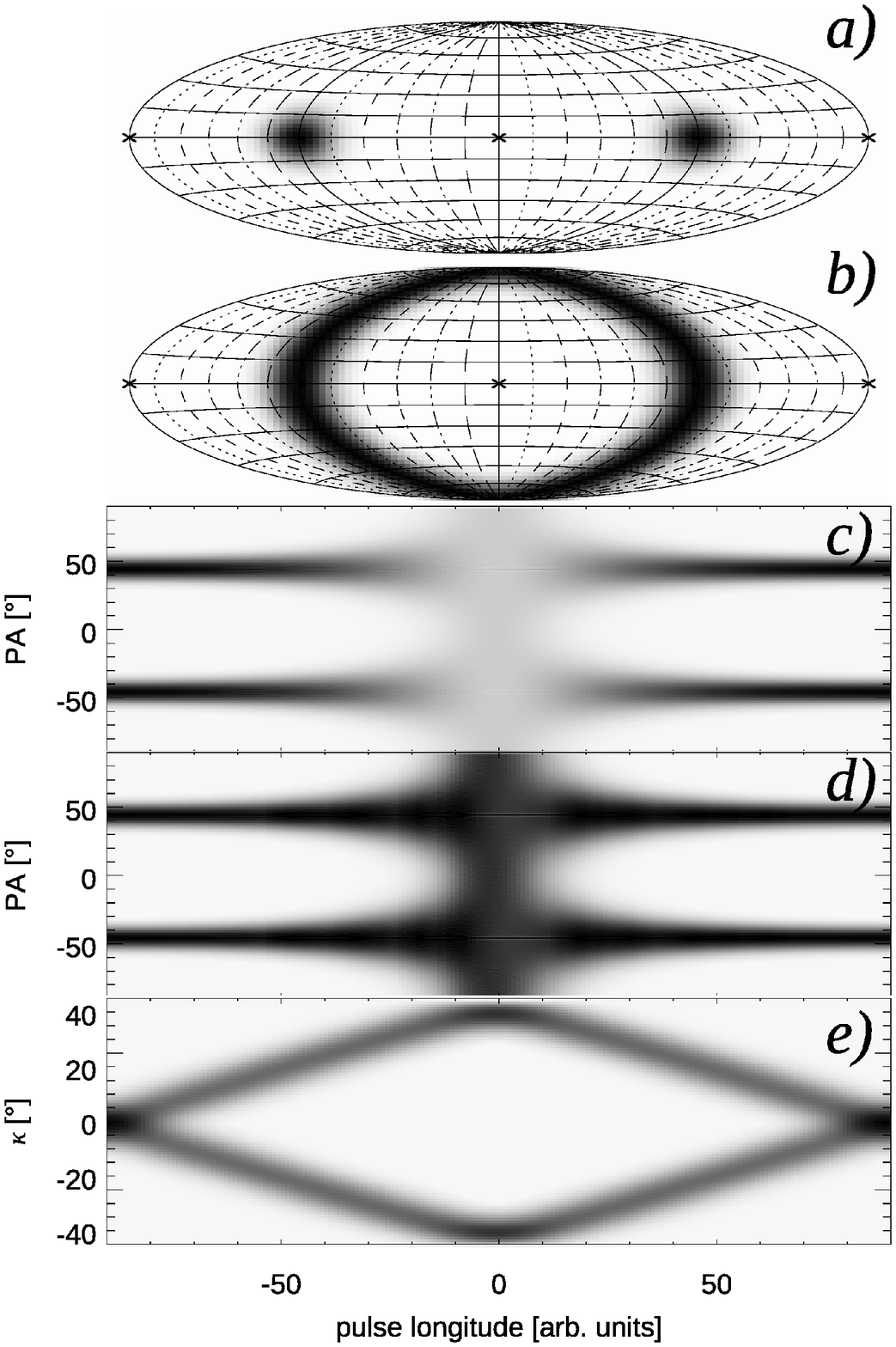}
\caption{Same as in Fig.~\ref{fivec} but for $\eta=0$ 
ie.~the patches move along a meridian of Poincar\'e sphere 
and pass through the V poles centrally. This leads to a pale break in the PA
curve (panel c) or a dark vertical column (d).
}
\label{fiveb}
\end{figure}

\section{V pole passage in the presence of two antipodal patches on the P.~sphere}
\label{numerical}

\subsection{Patches of equal strength}

As described in D20, the direction of OPM transition (up versus down)
depends on the side on which the flux patch is passing near the V pole on
the Poincar\'e sphere. Indeed, eqs.~(\ref{paone}) and
(\ref{patwo}) respectively contain factors $\tan\epsilon$ and $\sin\eta$, which
change sign when $\epsilon$ or $\eta$ does.   
In the presence of two imprecisely orthogonal polarization patches of equal
strength this leads to several possible transition geometries. 
 All the below-discussed cases correspond to the near-quarter-wave mode-ratio-driven coherent OPM
transition, ie.~$\dox=90^\circ-\eta\sim90^\circ$ and $\epsilon=0$. The patches are orthogonal
(antipodal) only initially since they move along slightly different great circles.

 Fig.~\ref{five} shows the case of two initially orthogonal patches 
rotating about two different axes slightly inclined with respect to the 
equatorial plane ($\eta=\pm7^\circ$). 
Both patches are passing on the same 
side of each V pole, ie.~both pass through the half-meridian zero near the poles,
and none through the half-meridian of $180^\circ$. 
The patches are initially located at the azimuth $\pm90^\circ$ in the
equatorial plane (Fig.~\ref{five}a) and are subsequently being rotated by $180^\circ$ around their
axes as marked with arrows in panel a). 
The left patch rotates about the axis which is cutting the P.~sphere at
points marked with `o' signs in panels a) and b). The right patch rotates about the axis
crossing the `x' points. 
The upward motion of the left patch, and the downward motion
of the right patch leave the round trace shown in Fig.~\ref{five}b. 
Tied by their near-orthogonality, the patches move 
in opposite directions near the V poles such that one patch crosses
 half-meridians with increasing values of azimuth, 
while the other crosses the same half-meridians in the direction of  
decreasing azimuth. Hence the OPM transition takes on the `X' form
 visible in 
Fig.~\ref{five}c and d. The numerical result is consistent with the analytical
solution of eq.~(\ref{pagen}) (solid lines). A more clear view of the
same numerical result (without the analytical solution overplotted), is
reproduced in Fig.~16 of Appendix.   

As described in D20, the immersion
of V pole within the flux patch spreads the flux across all PAs, hence the
'X' structure in panel c) becomes weak (the power is spread verticaly in the
longitude-PA plot). The same `X' feature is shown in d), albeit with the
greyscale normalization performed for each longitude separately, ie.~the black
color in d) corresponds to the maximum at a given longitude.
This highlights the X feature. 

The X form of the OPM transition resembles that observed at longitude
$172^\circ$ in PSR B1451$-$68 (between intervals A and B in
Fig.~\ref{data1}).  
The bottom panel (e) in Fig.~\ref{five} shows the ellipticity angle which corresponds to the
left half of the bow tie in Fig.~\ref{data1}. The numerical result follows
the analytical solution of eq.~(\ref{eligen}) (white lines). 

In Fig.~\ref{fivec} both patches are always orthogonal since
they are rotated about the same axis that 
cuts the P.~sphere at the `x' marks ($\eta=7^\circ$).  
Therefore, they pass the V poles on different sides (compare the near-polar trace of
upper and lower patch motion in panel b). However, since they also traverse
the azimuth in opposite directions (rightward for upper patch, leftward for
lower patch) the PA curve is bending in the same direction (downwards) in each PA track 
(see panels c and d). This type of the OPM transition has the S-shaped form,
similar to the well known S-swing of PA as caused by the unrelated RVM
effect.
As in the case of the X feature, each S-swing in Fig.~\ref{fivec}c
(and \ref{fivec}d) is described by eq.~(\ref{pagen}). 
As before, the ellipticity angle
of Fig.~\ref{fivec}e follows eq.~(\ref{eligen}). A version of 
Fig.~\ref{fivec} without the analytical lines overplotted is shown in Fig.~17 of
Appendix.

Finally, Fig.~\ref{fiveb} shows the case where both patches stay  
perfectly orthogonal and pass through their V
poles centrally. The patches are initially located at azimuths $\pm90^\circ$
(Fig.~\ref{fiveb}a) and move purely meridionally. In the course of this rotation, 
each PA track spreads equally strongly up and down, forming
a vertical column of weak radiative power, which looks like a
break in each PA track.\footnote{The figure-eight pattern that seems to be ghostly
formed in the PA track break of Fig.~\ref{fiveb}c, must be an optical
illusion, or some plotting artifact.} This type of OPM transition seems to
roughly 
correspond to the pulse longitude of $182^\circ$ in Fig.~\ref{data1} (between
intervals B and C).

In addition to these basic types of OPM transition, mixed cases are
possible, eg.~with one PA track  following the S curve while the other one
follows the break as expected for central pole passage.

\subsection{Patches of unequal strength}

\begin{figure}
\includegraphics[width=0.48\textwidth]{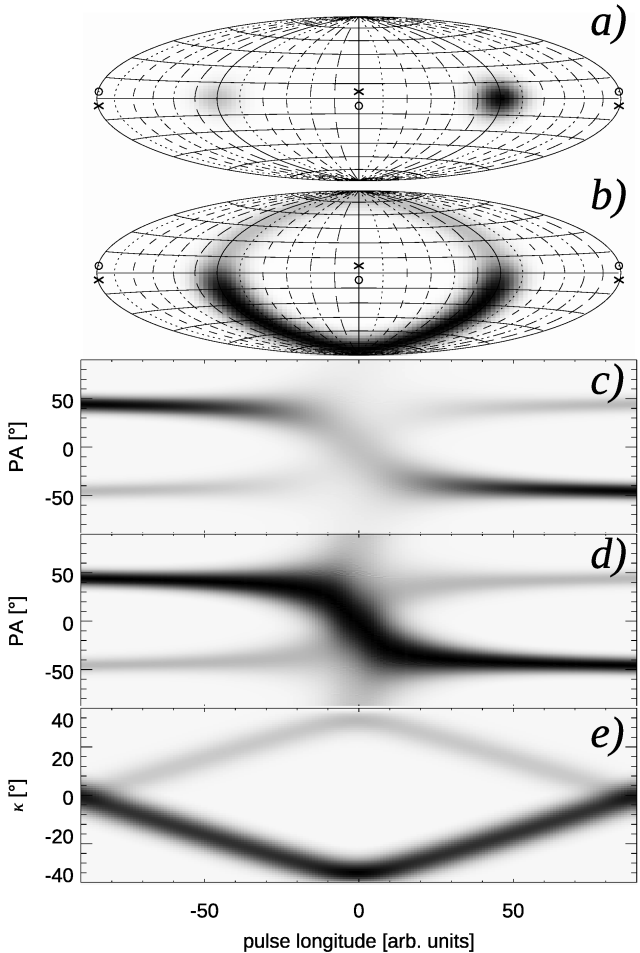}
\caption{Same as in Fig.~\ref{five} but with the initially left patch five times weaker
than the right hand side patch. The patches move around two different axes
with $\eta=\pm7^\circ$ 
so they pass on the same side of the V axis (as in the case of an OPM of the X type).
}
\label{uneq}
\end{figure}

\begin{figure}
\includegraphics[width=0.48\textwidth]{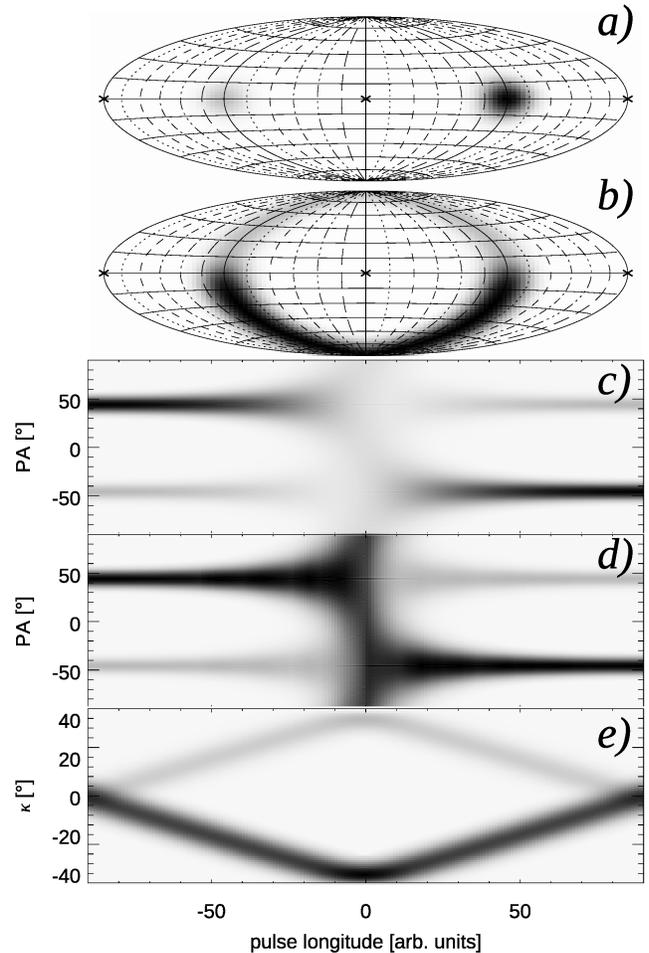}
\caption{Same as in Fig.~\ref{fiveb} but with a five times weaker initially left patch.
 In this figure $\eta=0$ which corresponds to the} central V pole passage. 
\label{uneqb}
\end{figure}

Figs.~\ref{uneq} and \ref{uneqb} present the pole-passage-induced OPM transition 
in the case of unequal modal power. Both modal patches are Gaussians of
$1\sigma$ width equal to $10^\circ$, but the peak flux in the secondary mode patch (initially left) is at 20\% of the other (initially right) patch. 

Fig.~\ref{uneq} shows the X type OPM transition which is similar to the case
of Fig.~\ref{five}, except that two 'arms' of the X feature are essentially invisible
(panel c) because of the weakness of the secondary mode and the vertical spread of
flux across all PA values. Fig.~\ref{uneqb} presents the case of central
pole passage with the same modal power ratio (20\%).


\section{Geometric model for the OPM power ratio inversion between the
leading and trailing profile side }
\label{leadtrail}

\begin{figure}
\includegraphics[width=0.48\textwidth]{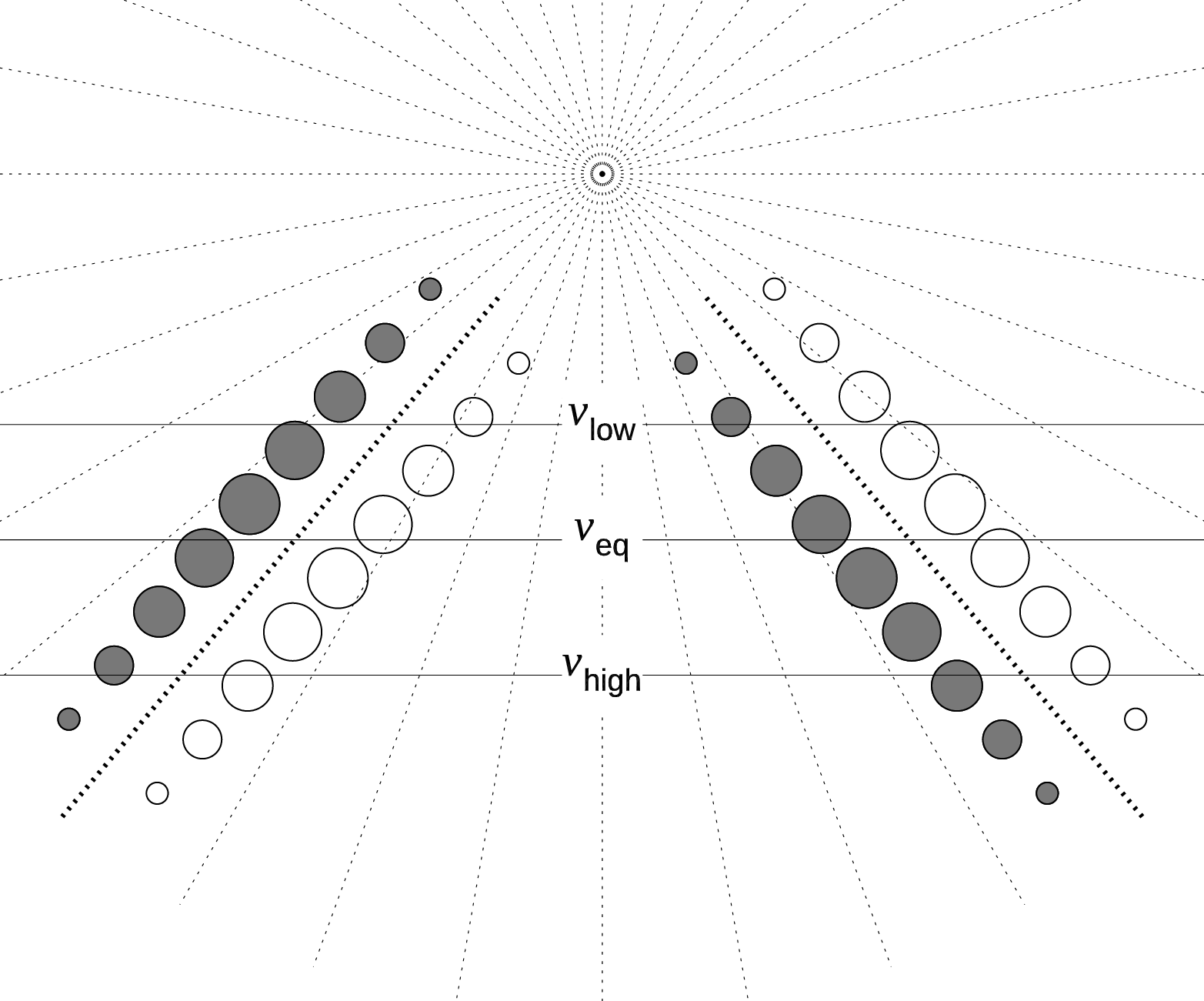}
\caption{Cartoon picture of sky-projected pulsar radio emission.
Strings of circles present near-instantaneous emission beams from charges
streaming away from the dipole axis (central point of the radial pattern of dots near the figure top). The
circles' diameter reflects the strength of emission, whereas different colors
correspond to different orthogonal polarization modes. Horizontal lines are paths of the
sightline corresponding to different observation frequency (the paths are
different to mimic the $\nu$-dependent size of the beam).  Both OPMs are comparable in brightness at $\nu_{eq}$. 
 }
\label{imbal}
\end{figure}

The polarization observed in the central part of profile (section B) in PSR
B1451$-$68 closely resembles the expected polarization of the vacuum curvature
radiation (CR). As illustrated in Michel (1991), \nct{m91} the CR beam is nearly
fully circular in the wings, whereas it is linear in the
plane of the curved particle trajectory. 
Indeed the low degree of circular polarization in the centre of the profile
($\Phi=178^\circ$) is not because individual samples of opposite sign of V are averaged out, 
as the green curve in panel b) of Fig. \ref{data1} demonstrates.
 Accordingly, while passing through
the CR beam, an observer will record the polarization state which moves from
one V pole to the opposite V pole (the sides of the CR beam have V of opposite
handedness). This would correspond to one tilted bar (eg.:~/) in the central X-shaped 
part of the observed bow tie.\footnote{The central X part of the bow tie is
formed by the ellipticity angle observed within 
the B interval, and should not 
be mistaken with the X feature of PA observed at the A/B 
transition.}
If the middle of the vacuum CR beam is ignored, however, 
then the fully circularly polarized wings correspond to two opposite polarization states (+V and
$-$V on the Poincar\'e sphere).

In one empirical model of pulsar polarization (D19), \nct{d2019} the two 
linear orthogonal modes (opposite patches on
the Poincar\'e sphere) are formed by incidence of a circular signal on
a linearly polarizing filter (as in a quarter-wave plate). To produce two different linear OPMs
with uncorellated amplitude, it was necessary to employ two circular signals of opposite
handedness.  
The signal needed in such model is then provided by the wings of the vacuum CR beam. 

Let us then assume that the emitted radio beam consists of two equally strong orthogonal modes
that have their emission directions separated by a small angle relative to the electron trajectory plane. 
In the fan beam model of pulsar emission, the quasi-instantaneous sky-projected emission of such
radiation could produce the pattern shown in Fig.~\ref{imbal}. The strings of circles
present radiation from two narrow streams of plasma flowing along
magnetic field lines. The streams emerge from
the vicinity of the dipole axis (near top of figure) and their projection on
the sky is shown with the thick dotted lines. Emission of each stream
involves two opposite polarization modes, illustrated with the grey and white
circles. Their transverse separation (across the stream) reflects the finite angular size of emission
beam. Local emissivity increases with the circle's size, thus the
intensity first increases with the distance $\theta_m$ from the dipole axis,
then falls off. 
Because of this intensity gradient and the oblique cut through
the streams, the line of sight is usually sampling unequal amounts of two OPMs
(compare the size of the grey and white circles cut by the bottom path of the line of sight). 
However, 
since the tilt of the streams is opposite on the leading and trailing side, 
different modes dominate in the leading and trailing side of the profile.
The ratio of OPMs depends on the angular scale of the emissivity gradient and the obliqueness of the cut. Likely effects of averaging caused by the spatial
extent of the emitter are ignored here for simplicity. However, it
should be noted that a uniform ring (cone) would not produce any modal
imbalance -- the effect is present only in the fan beam model (in the conal model
the emissivity would have to be azimuthally nonuniform (on average) for the mode imbalance
to appear in average profiles). The magnitude
of mode imbalance thus depends on how nonunifrom the emissivity is 
in both the magnetic azimuth (measured around the magnetic dipole axis) and in 
the magnetic colatitude. 

Because of the spatial convolution effects, 
it may be impossible to directly resolve the bifurcated bimodal emission beam
of Fig.~\ref{imbal}. Even in the presence of a convolution, however,
 such a beam can produce the observed OPM exchange, ie.~the dominance
of different OPMs in different sides of profiles (leading vs trailing). If the emissivity is
strongly nonuniform in magnetic azimuth, the mode dominance can change
several times within the profile (which is observed eg.~in J0437$-$4715,
Oslowski et al.~2014). \nct{ovb14}

Note that the scenario of Fig.~\ref{imbal} refers to the OPM ratio inversion
observed both in the patches and in the PA tracks, ie.~between the leading and trailing edge of the
profile. The COMT-related V-pole passage requires additional OPM transitions in which the flow of radiative
power occurs between the orthogonal PA tracks, but not between the patches.  
These two types of OPM transitions coexist in the profile and they contribute to the
complexity of PA tracks on the longitude PA diagrams (see D20).

\subsection{Frequency dependence of real OPM power exchange}

Intrinsic emission of two modes with similar strength is unlikely, because
of their different amplification in anisotropic coherent processes (Melrose
2003). \nct{m03}
The model of Fig.~\ref{imbal} provides a geometric interpretation for a 
small imbalance of the modal strength in radio pulsar signal, under the
assumption that equal amount of each mode is emitted in different parts of
the radio beam. Comparable power in different OPMs is
often observed, and sometimes one mode dominates at low $\nu$ whereas the other at high
$\nu$ (Noutsos et al.~2015; Young \& Rankin 2012). \nct{yr12, nsk15}
In such case there exists a frequency $\nu_\mathrm{eq}$ at which both modes are
perfectly equal. Strong evolution of PA tracks is
observed near this transition frequency (Navarro et al.~1997;  Mitra et al.~2016). 
\nct{nms97, mra2016} 

These phenomena can be understood with the model of Fig.~\ref{imbal},
assuming that the beam projections follow radius to frequency mapping,
ie.~they move away from the dipole axis at low $\nu$. 
Instead of three different beam positions, three different paths of
sightline are marked on the same beam in Fig.~\ref{imbal} (solid horizontal
lines). At high $\nu$ the line of sight is traversing through the beam
periphery (bottom horizontal line), so the `white mode' is stronger on the
leading side, whereas the `grey mode' on the trailing side (compare the
size of white and grey circles). At low $\nu$ (top horizontal) the grey mode is brighter on
the leading side (and the white on the trailing one). At the moderate frequency
of $\nu_\mathrm{eq}$ equal
amounts of modes are observed (though not necessarily in all profile components
at the same $\nu_\mathrm{eq}$). Since the radiation beam becomes narrower at higher
$\nu$ the spatial convolution could depolarize the profile at 
higher frequencies, at least in the regions with an oblique cut through the
streams.  

 Fig.~\ref{imbal} implies that the ratio of OPMs 
is changing with $\Phi$ and $\nu$. However, the figure predicts an essentially fixed
profile width. The observed low-$\nu$ profile widening (Wu et al.~1998) \nct{wgr98} 
must be related to the effect
that causes the below-discussed interference phenomenon. 

\section{Implications for physics: direct beam or coherent superposition of adjacent components?}
\label{physical}

\subsection{Direct CR microbeam}

The central part of the bow tie (section B in Fig. \ref{data1}) is very similar to the polarization
observed while crossing the vacuum curvature radiation beam: 
the polarization state travels from one V pole to the other (opposite) V pole. 
For example, a possible sequence of three polarization states is: $+V, +Q, -V$ 
(with all the unspecified Stokes components equal to zero).
This is equivalent to the motion along half a meridian of the P.~sphere.  
However, the polarization state observed within the full pulse window of
PSR B1451$-$68 traces
almost a full meridional circle. 
So the previous sequence would be extended to: 
$-Q, +V, +Q, -V, -Q$. 
Thus the polarization becomes linear in the
profile wings. It is not clear whether this extra rotation of the polarization
state (towards the equator of P.~sphere) 
in the profile wings is intrinsic or results from spatial convolution or propagation
effects. Definitely, the unmodified vacuum CR microbeam is not consistent with the 
polarization observed in PSR B1451$-$68.


\subsection{Coherent superposition of adjacent components}
\label{overlap}

The three regions of high linear polarization (A, B, and C in
Fig.~\ref{data1}) may correspond to the three distinct profile components
observed at lower frequencies ($171$ and $271$ MHz, Wu, Gao , Rankin et
al.~1998). \nct{wgr98} Dips (depressions) in total intensity $I$ on both sides of the core
component (at $\Phi=170^\circ$ and $184^\circ$ in Fig.~\ref{data1}a) are also 
consistent with the triple profile form. 
Assuming that the adjacent components have orthogonal linear
polarization (see Fig.~\ref{modsep}b), the coherent superposition of radiation within the regions where
they overlap could produce the meridional circularization. The tips of the bow tie 
(at $\kappa\approx\pm45^\circ$) would then correspond to 
longitudes where intensities of the modes in adjacent components are
equal (to get the pure circular polarization). 
This
interpretation is consistent with the coincidence of dips in the profile with
the bow tie tips (at $\kappa=\pm45^\circ$). Such scenario, however, employs
a longitude-dependent mode brightness ratio and requires that radiation in the
adjacent components maintains fixed phase lag difference of $\delta=90^\circ$ 
(or at least that $\delta\approx90^\circ$ when the components contribute equally). 
More specifically: the radiation in a given polarization mode in component A 
must be coherently summed with orthogonal mode in component B, which must be oscillating at phase lag
$\delta=90^\circ$ (as measured with respect to the oscillation in component A).  
Moreover, for the polarization state to continue its circulation in the same
direction, the next profile component (in region C) should keep a phase lag
of $180^\circ$ with respect to component A. Thus, to keep the unidirectional
rotation of the patch, the linearly polarized radiation in components A, B, C should have the 
oscillation phase increasing in steps of $90^\circ$ (eg.~$0$, $90^\circ$
and $180^\circ$ for components A, B, and C, respectively). 
It is then concluded that the monotonic increase of the oscillation phase lag with pulse longitude is 
needed in the model based on the variable mode ratio (as
determined by shapes and locations of overlapping profile components).  
 Apparently, the components are formed by radiation with well defined
oscillation phase, which increases monotonously.

\section{The model: interference pattern of four modes}
\label{sinterf}

\subsection{Pulsar radiation as the coherent superposition of four modes} 

Importantly, both the radiation in the central component
(`core') as well as radiation in the `conal' components consists of two
polarization modes. Therefore, in the longitude intervals where the
core and cone components overlap, four radiative contributions are superposed: the O and X mode of the core, 
and the conal O and X modes.\footnote{In D19 the equal amounts
of OPMs were attributed to `linearly-polarizing filtering' of an initially 
circularly-polarized signal. Four linear modes have appeared necessary in such
model to account for opposite handedness and uncorellated strength of observed
OPMs (Sect.~3.2 and fig.~7 therein). Recently, four modes have been invoked for the modulated 
polarization of B1919$+$21, W. van Straten, private communication.} 
When the O mode is coherently superposed with the X mode, the strong circular polarization
appears. However, the coherent summation of the two O modes (from core and
cone), only affects the intensity of the total O mode (and similarly for the two X
modes). Depending on the value of the
phase lag $\delta_{oo}$ between O$_\mathrm{core}$ and O$_\mathrm{cone}$, the total intensity will be enhanced
or reduced. The phase lag will likely depend on frequency $\nu$ and 
pulse longitude $\Phi$.

If, at some frequency $\nu_1$, the value of $\delta_{oo}(\nu_1)$ (between O$_\mathrm{core}$ and
O$_\mathrm{cone}$, both oscillating in the same plane) 
is a multiplicity of $360^\circ$, the 
total intensity will be stronger (positive interference). 
If the phase lag $\delta_{ox}$ between the total O mode and the remaining X mode is close
to $\sim90^\circ$, their coherent superposition will produce the observed rotations of the polarization state at both
sides of the core (circularization). If $\delta_{ox}\sim0$ the PA of total
signal will have the PA deflected by $45^\circ$ away from the O and X modes (a $45^\circ$ jump, D19). 
However, if there is a distribution of phase
lags $\delta_{ox}$ between the (total) O and X
mode, the observed signal will become weakly polarized (instead of
circularly polarized).  
On the other hand, if, at some different frequency $\nu_2$,
$\delta_{oo}(\nu_2)\sim180^\circ$, then 
the two O waves will cancel each other (negative interference). This will decrease $I$ and will produce the deep
minima between the components. The flux at such minima will be
dominated by the other mode (X). 
This must be the reason for the ubiquitous
enhancement of polarization degree at minima in many pulsar profiles, eg.
in B1929$+$10 (Rankin \& Rathnasree 1997) \nct{rr97} 
and in the central parts of several D type profiles (as defined in
Rankin 1983). \nct{ran83}  
This four mode model thus resolves the long-standing issue of why
 the `superposition of two overlapping components' does not 
depolarize the minimum between
the `components': the minima appear because one mode is suppressed, hence the
polarized fraction increases. 

In such way both the circularization and the minima between the
components are produced by the same effect: the coherent (or partially
coherent to incoherent) summation of
orthogonal and parallel modes in radiation from two superposed signals 
(eg.~a direct beam and a reflected beam, somewhat similar to lasers, though
here each beam contains two polarizations).
It is then concluded that the well separated components in the triple-form
low-$\nu$ profile of B1451$-$68 may well have essentially the same origin as the  
circularization: it is the coherent summation of oscillations that are either
orthogonal or parallel to each other (in various possible combinations). 

\subsection{The other way round: components from superposition, not
 superposition of components}

It has been found above that different components in the average profile of B1451$-$68
consist of radiation with a well defined oscillation phase (or phase lag),
contrary to the mixture of phase lags which is normally expected in
astrophysical sources. 
This macroscopic (profile-wide) structure of phase lags, along with the
clearly demonstrated superposition of four modes, strongly suggest that 
pulsar profiles are mostly (or at least partially) an interference pattern: 
 they must involve two radiative signals, each consisting of two orthogonal
modes. The shape of profile (hence modulation properties) result from positive or
negative interference of these two signals. 

This means that the minima between `components' in pulsar profiles are just
regions of negative interference ($\delta_{oo}$ or $\delta_{xx}$ being roughly
$180^\circ$),  
so the `components' (at least some of them) do not actually exist as
separate entities -- they just present the
points in space-time where the interference pattern has maxima sampled by 
the line of sight (as determined by the usual special-relativistic
detectability conditions). In other words, the observed pulsar signals
represent a cut through the interference pattern, whereas the spatial
emissivity distribution (ie.~shape of emission region) 
 contributes to the profile shape, 
 but it is not the only factor.  

The model based on an interference pattern has a major advantage over a model
which is plasma-density-governed. Given the approximate mirror symmetry of the average
profile and the axial symmetry of the open field line region, a
time-symmetric refraction index and phase lag seem most natural. 
In the case of interference pattern, the longitude (or time) dependence of the phase lag does not have to
be symmetric about the profile center. The polarization state and its associated phase lag is determined 
by the space-time structure of the interference pattern. 

The physical cause of this
interference, which occurs within the radio pulsar beam, remains to be established.
Possible processes that lead to interference include reflection (or
scattering), refraction (Edwards et al.~2003) \nct{esv03} and 
emission from two sources, eg.~emission directed in opposite directions along field lines. 
It is also
possible that the four modes originate from the linearly-polarizing filtering of 
circularly polarized signals (D19) or elliptically polarized
signals, as provided by the curvature radiation beam (Fig.~\ref{imbal}). The
conditional equivalence of two elliptical modes to the four wave model is
discussed below.\footnote{A three-mode model, with interference of two O
mode signals (supplemented with the X mode), or with interference of two X modes
(plus O mode), has similar properties to the four-mode model.} 

\begin{figure}
\includegraphics[width=0.48\textwidth]{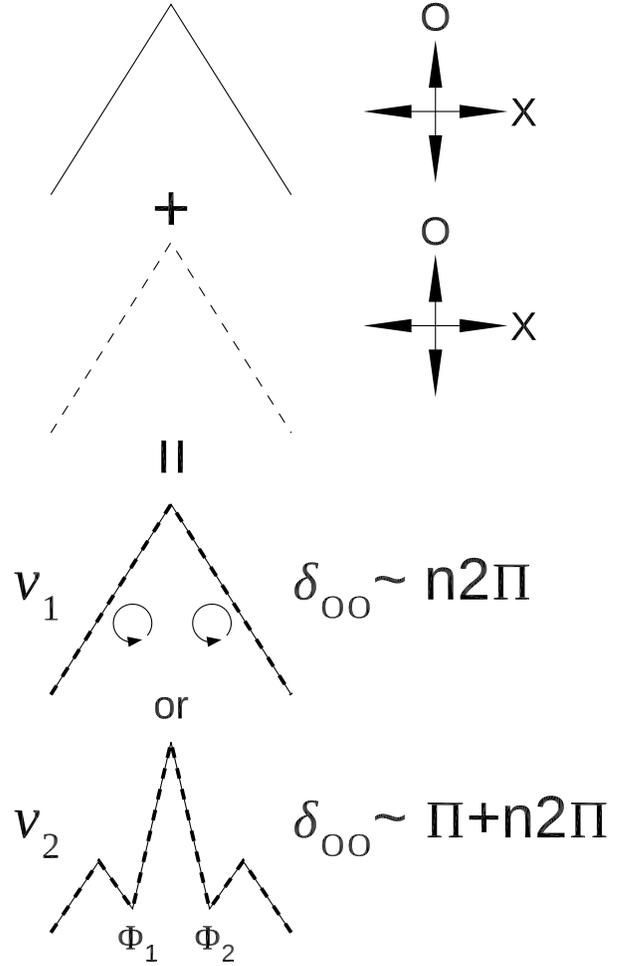}
\caption{Mechanism of profile shaping by coherent superposition of two triangular profiles
shown in top part of figure. Each triangle contains two orthogonal modes shown on the right. 
Depending on the frequency-dependent phase lag
$\delta_{oo}$ between the two coplanar O modes, 
the total profile may reveal the minima (and triple form) or not. For
positive interference at $\nu_1$ the additional contribution of X mode can produce the
circularization (if $\delta_{ox}\sim90^\circ$). At $\nu_2$ the two O modes
cancel each other, so the presence of the X mode increases the polarization fraction
at the minima (not shown). The values of $\delta_{oo}$ 
only refer to longitudes $\Phi_1$ and $\Phi_2$, and interference of the two
X modes is ignored for simplicity.
 }
\label{interf}
\end{figure}

By applying the four mode model for PSR B1451$-$68, the profile can even be understood as a
superposition of two triangular profiles which peak at the same 
longitude (Fig.~\ref{interf}). The deep minima observed on both
sides of the central peak (`core') at low $\nu$ 
result from the negative interference, as determined
by the longitude dependence of the oscillation phase (phase lag). The
longitude structure of the oscillation phase, on the
other hand, is intrinsic to the interference pattern. The frequency-dependent
position of minima, eg.~their increasing distance at low $\nu$ must be attributed in such model to the
wavelength-dependence of the interference pattern.  

The average profiles are known to be averaged modulation patterns. 
Such superposition model should thus be extended to observed time modulation effects
such as drifting subpulses. An outstanding example of this phenomenon is the structure of 13
subpulses observed in PSR B0826$-$34 (Esamdin et al.~2005). \nct{esa2005} 
According to the four-mode interference model, the subpulses of B0826$-$34 mostly originate
from the phase lag structure of the superposed beams, and not from the beam
pattern alone (as in the carousel model). 
The similar widths and separations of subpulses in such scenario thus 
correspond to distances in the interference pattern. 

Related phenomena are the changes of modulation or subpulse drift pattern (and the associated
profile modes) see, eg.~fig.1 in Weltevrede (2006). These can be interpreted as
disturbances, destruction or destabilization of the interference pattern. 
 The interference of radiative signals is a different process than
the oscillations within the emission region itself, as considered by Clemens \& Rosen
(2004). In particular, only the superposition of precisely orthogonal modes has 
been considered in Clemens \& Rosen (2008) \nct{cr2004, cr2008} 
with no coplanar interference
possible. In the present model the anticorellation between 
OPMs is produced through the interference of radio waves within the pulsar radio
beam. 

It is concluded that the so far mostly ignored necessity to take into account the
coherent effects of phase lags  
is the key reason for several long-standing difficulties
with the carousel model (McSweeney 2019; McSweeney et al.~2019; Maan 2019).
\nct{sweephd2019, swee2019,
maa2019}
The negative interference of
radiative contributions oscillating in the same plane creates the illusion of components
considered as separate entities. 
Instead, it seems there is
no rotating carousel there at all - rather a different emission
region with the outflowing interference pattern and the geometry of sightline sampling.

\subsection{Four linear modes versus two elliptic modes}

To explain the similar strength of OPMs, it has been assumed in D19 that the initial
signal is circularly polarized, and it is passing through a magnetospheric region which transmits
only two orthogonal linear modes (linearly birefringent filter). To finally obtain
the observed 
two elliptical OPMs of opposite handedness, however, two circular signals of
opposite handedness were needed, each one producing a pair of linear OPMs (see fig.~7 therein).
 
A pair of orthogonal linear modes in general corresponds to an elliptically polarized
signal, hence the four linear modes can be considered as just two elliptical
signals. The final conclusion of D19, therefore, was that a pulsar signal
originates from coherent superposition of two elliptically polarized modes. 
Mathematically, it is anyway convenient to express the two elliptical modes as four
linear waves. Equations (11)-(14) in D19 describe such four oscillations. 
If arbitrary amplitudes of the elliptical waves are allowed, the two waves can be
written as: 
\begin{eqnarray}
E_1^x&=&A_1 \cos\kappa_1 \cos(\omega t-kz)\label{wone}\\
E_1^y&=&A_1 \sin\kappa_1 \sin(\omega t-kz)
\end{eqnarray}
\begin{eqnarray}
E_2^x&=&A_2 \sin\kappa_2 \sin(\omega t - kz - \lag)\\
E_2^y&=&A_2 \cos\kappa_2 \cos(\omega t - -kz -\lag)\label{wfour}
\end{eqnarray}
where $\kappa_1$ and $\kappa_2$ are the ellipticity angle of the two
superposed waves  (the angles were denoted $\beta$ in D19). 
If $\kappa_1 = \kappa_2 = \kappa$, the elliptical waves are
orthogonal, ie.~their Stokes vectors are antipodal on the
P.~sphere.\footnote{This can be shown by 
calculating the polarization state vectors for the separate elliptical waves, 
ie.~$\vec S_1=(Q_1, U_1, V_1)$ and $\vec S_2=(Q_2, U_2, V_2)$. Then 
$\vec S_1\times\vec S_2 = 0$ and the corresponding components have opposite signs,
ie.~$\vec S_1$ and $\vec S_2$ have opposite directions.}
Note that one phase lag $\lag$ between the two elliptical waves is sufficient to
parametrize the phase relations in this two waves model (the lag between the
$x$ and $y$ components is fixed to $\pm 90^\circ$).

The Stokes parameters for the coherent sum of these waves, (ie.~$E_1^x +
E_2^x$ and $E_1^y+E_2^y$) can be calculated as:
\begin{eqnarray}
I & = & A_1^2 + A_2^2 - \nonumber\\ 
& &  A_1 A_2 \sin\lag (\cos\kappa_1\sin\kappa_2-\sin\kappa_1
\cos\kappa_2)\label{intjed}\\
Q & = & A_1^2 (\cos^2\kappa_1 - \sin^2\kappa_1) + A_2^2(\sin^2\kappa_2 -
\cos^2\kappa_2)\nonumber\\ & & 
- 2 A_1 A_2\sin\lag[\cos\kappa_1\sin\kappa_2 + \sin\kappa_1\cos\kappa_2]\\
U & = & 2 A_a A_2 \cos\lag (\cos\kappa_1\cos\kappa_2 + \sin
\kappa_1\sin\kappa_2) \\
V & = & 2[A_2^2\sin\kappa_2\cos\kappa_2 - A_1^2\sin\kappa_1\cos\kappa_1 + \nonumber\\
  & & A_1 A_2 \sin\lag (\sin\kappa_1\sin\kappa_2 -
\cos\kappa_1\cos\kappa_2)]\label{vjed}
\end{eqnarray}
The lag-dependent term in (\ref{intjed}) corresponds to positive or
negative interference which is possible if the elliptical waves are not
orthogonally polarized. Indeed, for orthogonal modes $\kappa_1=\kappa_2=\kappa$ the Stokes parameters become:
\begin{eqnarray}
I & = & A_1^2 + A_2^2\label{intdwa}\\
Q & = & (A_1^2 - A_2^2) \cos(2\kappa) 
- 2 A_1 A_2\sin\lag\sin(2\kappa)\\
U & = & 2 A_1 A_2 \cos\lag \\
V & = & \sin(2\kappa)(A_2^2-A_1^2)  
- 2 A_1 A_2 \sin\lag \cos(2\kappa)
\end{eqnarray}
As can be seen, the interference term disappears from eq.~(\ref{intdwa}),
and the result is the same as in an incoherent case (sum of amplitude squares 
instead of the square of the sum). This is caused by the independence
(orthogonality) of the elliptical modes.

It is then concluded that the model which employs four arbitrary but
linearly-orthogonal waves is not equivalent to the model
based on the sum of two elliptically-orthogonal waves (modes). They become the same only if the two
pairs of waves combine to two elliptically-orthogonal waves, as represented
by two antipodal
 states on the P.~sphere. In such case the interference does not
occur (despite the coherent superposition). This is the reason why the
interference term is present in eq.~(\ref{intjed}) but it is missing in
eq.~(\ref{intdwa}).\footnote{For this same reason, there is an interference
term in eq.~(24) of D19, but no such term in eq.~(28) of that paper.}

\def\ejx{{\cal E}_1^x}
\def\ejy{{\cal E}_1^y}
\def\edx{{\cal E}_2^x}
\def\edy{{\cal E}_2^y}

The intensity-affecting interference can thus occur between imprecisely 
orthogonal elliptical modes. The elliptical modes in eqs.~(\ref{intjed})-(\ref{vjed}) 
differ in $\kappa$ so the modes are deflected from orthogonality in the meridional plane 
on the P. sphere. 
Interestingly, there are observations in which the modes are
misaligned approximately in this way (eg.~B0031$-$07, fig.~5 in Ilie et
al.~2020).
\nct{iwj19}

The above described model involves only five parameters: two amplitudes, two
ellipticities, and the phase lag. 
A more general model assumes the superposition of four independent
linearly-polarized waves of arbitrary origin, 
with four arbitrary amplitudes and four oscillation phases:
\begin{eqnarray}
E_1^x&=&\ejx\cos(\omega t-kz-\phi_{1x})\\
E_1^y&=&\ejy\cos(\omega t-kz-\phi_{1y})
\end{eqnarray}
\begin{eqnarray}
E_2^x&=&\edx\cos(\omega t - kz - \phi_{2x})\\
E_2^y&=&\edy\cos(\omega t - kz - \phi_{2y})
\end{eqnarray}
and the corresponding Stokes parameters, calculated in the usual way
(D19), are:
\begin{eqnarray}
I&=&(\ejx)^2 + 2\ejx\edx\cos{(\phi_{1x}-\phi_{2x}) + (\edx)^2}\nonumber \\
& & +  (\ejy)^2 + 2\ejy\edy\cos{(\phi_{1y}-\phi_{2y}) + (\edy)^2}\label{ist}\\
Q&=&(\ejx)^2 + 2\ejx\edx\cos{(\phi_{1x}-\phi_{2x}) + (\edx)^2}\nonumber \\
& & - (\ejy)^2 - 2\ejy\edy\cos{(\phi_{1y}-\phi_{2y}) - (\edy)^2}\\
U&=&2[\ejx\ejy\cos(\phi_{1x}-\phi_{1y}) + \ejx\edy\cos(\phi_{1x}-\phi_{2y})
+
\nonumber \\
& & \ejy\edx\cos(\phi_{1y}-\phi_{2x}) + \edx\edy\cos(\phi_{2x}-\phi_{2y})] \\
V&=&2[\ejx\ejy\sin(\phi_{1x}-\phi_{1y}) + \ejx\edy\sin(\phi_{1x}-\phi_{2y})
+  \nonumber\\ 
& & \ejy\edx\sin(\phi_{2x}-\phi_{1y}) + \edx\edy\sin(\phi_{2x}-\phi_{2y})] \\
\nonumber
\end{eqnarray}
The cosine terms in eq.~(\ref{ist}) correspond to the negative or
positive interference. The waves are orthogonal (noninterfering) when
the $x$ oscilations in the two waves (and simultaneously the $y$ oscillations) 
are delayed by $90^\circ$ (plus multiplicity of $180^\circ$), ie.~when
$\cos(\phi_{1x}-\phi_{2x})=0$ and $\cos(\phi_{1y}-\phi_{2y})=0$. 
It is worth to note that even if an observed signal actually consists of two elliptical
signals (orthogonal or not), its polarization may be easier understood 
by considering the superposition of four linear waves instead (as done in the previous
subsections).

\section{Conclusions}

The polarization properties of PSR B1451$-$68 reveal that the radiative power
contained within orthogonal PA tracks can change because of two independent
processes: either by the near-meridional V pole passage of the polarization
state, or by the traditional change of the modal strength ratio. This
reemphasizes the need to follow the variations of pulsar polarization as a
function of pulse longitude on the Poincar\'e sphere, where the modes
(patches) can be more easily separated than on the standard longitude-PA
diagrams. 

 The meridional circularization is caused by the coherent 
OPM transitions at the fixed quarter-wave lag ($\dox=90^\circ+n180^\circ$). It is not caused
by the retardation (increasing $\dox$) at a fixed and equal mode amplitude. 
The coherent orthogonal mode transition is described by an equation of
similar form to the RVM equation. 
Since the polarization state simultaneously takes part in two motions: RVM
and VPP, the polarization angle as a function of pulse longitude follows a
curve which is the sum (eq.~\ref{pasum}) of two curves: the RVM (eq.~\ref{rvm})
and the RVM-like COMT curve (eq.~\ref{pagen}). 

The COMT may therefore be mistaken for
an RVM swing. In the quarter-wave case, the high circular polarization degree may help 
to discern between these two scenarios, 
however, in average profiles V may be suppressed 
by the noncoherent contribution of the other orthogonal mode. 
 In such case it is useful to plot the typical instantaneous polarized
fractions instead of their Stokes-based average. This COMT can also be recognized on the
P.~sphere.
In the case of a half-wave COMT ($\dox\sim n180^\circ$), 
the polarized fractions are not affected and 
the pol.~state moves within the QU equator so the PA changes gradually 
and may be impossible to discern from RVM. 

 We have shown that the profile components of PSR B1451$-$68 consist of radiation with
different and well defined oscillation phase. The phase increases
monotonously with pulse longitude 
and each component consists of two orthogonal polarization modes.  
Such properties imply that pulsar radio emission can be
understood as the superposition of two radiative contributions, each of which contains both
linearly polarized modes, ie.~the observed characteristics 
result from summation of four radiative contributions. This is mathematically equivalent to the
superposition of two non-orthogonal elliptically polarized waves. 
Therefore, the profile components (hence subpulses) are determined (or at least affected) by 
the space-time structure (hence longitude dependence) of the phase lag between the coherently added
parallel and orthogonal polarizations. To a large degree, then, the carousel
(and conal) picture of the pulsar emission region may be an illusion produced by the
cancelling or enhancement of coherently superposed signals.

\section*{acknowledgements}
We thank S. Johnston for help with the observations. 
We also thank Aris Karastergiou for reviewing this manuscript, leading to various improvements.
This work was supported by the grant 2017/25/B/ST9/00385 of the National Science
Centre, Poland. Pulsar  research  at  Jodrell Bank  Centre  for  Astrophysics  and  Jodrell  Bank  Observatory  
is supported  by  a  consolidated  grant  from  the  UK  Science  and Technology Facilities Council (STFC). 
The Parkes telescope is part of the Australia Telescope National Facility which is funded by the Commonwealth 
of Australia for operation as a National Facility managed by CSIRO (Commonwealth Scientific and  Industrial  
Research  Organisation).  

\section*{Data availability}

The data underlying this article can be accessed from the 
CSIRO Data Access Portal (https://data.csiro.au/dap), in the ATNF pulsar observation domain. 
The analysed dataset for PSR J1456-6843 has the unique identifier t160417$_-$201032.sf. 
The data used in the calibration process has the identifier t160417$_-$200715.sf. 
The derived data generated in this research will be shared on reasonable request 
to the corresponding author.

\bibliographystyle{mn2e}
\bibliography{listofrefs2}

\section*{Appendix}

\begin{figure}
\includegraphics[width=0.48\textwidth]{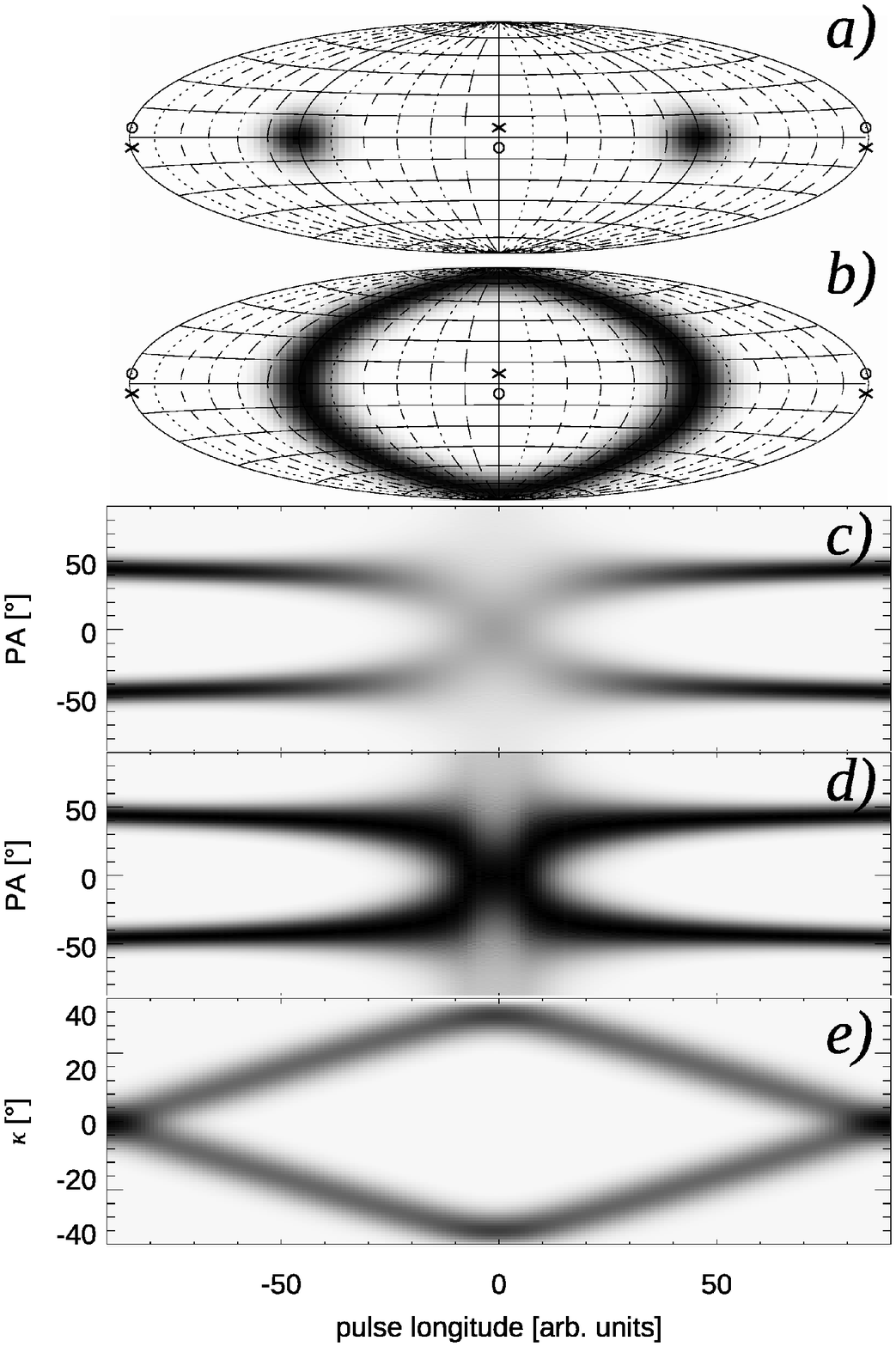}
\caption{Reproduction of Fig.~\ref{five} without the analytical solutions of
eqs.~(\ref{pagen}) and (\ref{eligen}).
}
\label{fpan}
\end{figure}

\begin{figure}
\includegraphics[width=0.48\textwidth]{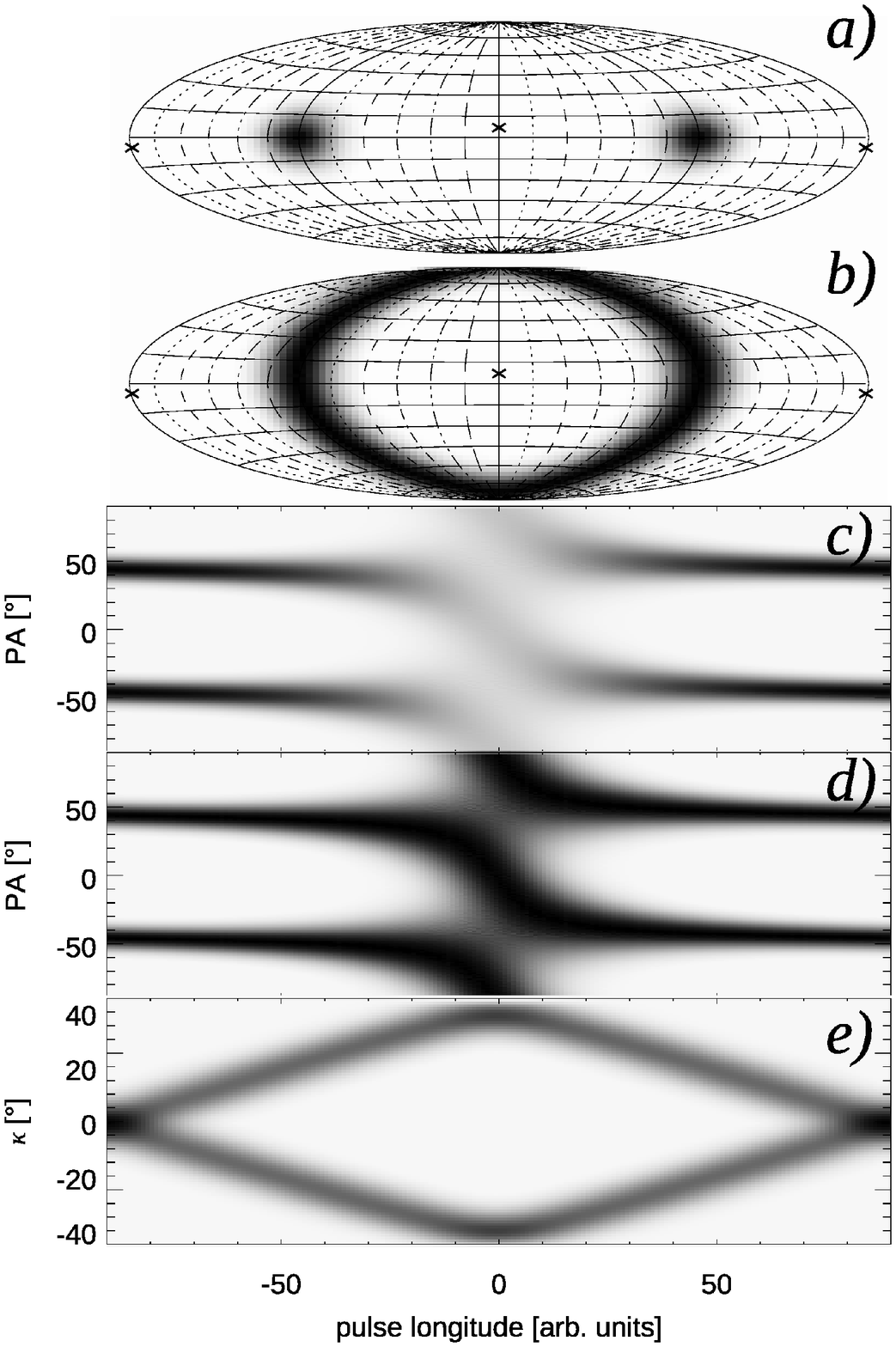}
\caption{Reproduction of Fig.~\ref{fivec} without the analytical solutions of
eqs.~(\ref{pagen}) and (\ref{eligen}).
}
\label{fpanc}
\end{figure}

The lines described by eqs.~(\ref{pagen}) and (\ref{eligen})  
make the view of the grey-scale patterns in Figs.~\ref{five} and \ref{fivec} less clear. Therefore, the figures
are reproduced here without these lines. 
Otherwise, Figs.~\ref{fpan} and \ref{fpanc} 
are identical to Figs.~\ref{five} and \ref{fivec},
respectively.


\end{document}